\documentclass[10pt,twocolumn,letterpaper]{article}
\usepackage[utf8]{inputenc}
\usepackage[margin=0.8in]{geometry}
\usepackage{setspace}
\usepackage{graphicx}
\usepackage{booktabs}
\usepackage{amsmath}%
\usepackage[labelfont=bf,skip=5pt,justification=justified, format=plain]{caption}
\usepackage[caption=false]{subfig}
\usepackage[style=chem-acs,articletitle=true,maxbibnames=30]{biblatex}
\usepackage [english]{babel}
\usepackage [autostyle, english = american]{csquotes}
\MakeOuterQuote{"}
\usepackage{lipsum}
\usepackage{url}
\usepackage{multirow}

\begin{document}

\begin{singlespace}

\title{La$_4$Co$_4$X (X = Pb, Bi, Sb): a demonstration of antagonistic pairs as a route to quasi-low dimensional ternary compounds}
\author{Tyler J. Slade$^{1*}$, Nao Furukawa$^{1,2}$, Matthew Dygert$^{1,3}$, Siham Mohamed,$^{1,4}$ Atreyee Das$^{1,2}$, \\ Weiyi Xia$^{1,2}$, Cai-Zhuang Wang$^{1,2}$, Sergey L. Bud’ko$^{1,2}$, Paul C. Canfield$^{1,2*}$}
\date{}

\twocolumn[
\begin{@twocolumnfalse}

\maketitle

\begin{center} 
    
\textit{$^{1}$Ames National Laboratory, US DOE, Iowa State University, Ames, Iowa 50011, USA} \\  
\textit{$^{2}$Department of Physics and Astronomy, Iowa State University, Ames, Iowa 50011, USA} \\
\textit{$^{3}$Department of Physics and Astronomy, University of Missouri, Columbia, Missouri 65211, USA} \\
\textit{$^{4}$Department of Chemistry, Iowa State University, Ames, Iowa 50011, USA} \\

\begin{abstract}

We outline how pairs of strongly immiscible elements, referred to here as antagonistic pairs, can be used to synthesize ternary compounds with low or quasi-reduced dimensional motifs intrinsically built into their crystal structures. By identifying third elements that are mutually compatible with a given antagonistic pair, ternary compounds can be formed in which the third element segregates the immiscible atoms into spatially separated substructures. Quasi-low dimensional structural units, such as sheets, chains, or clusters are a natural consequence of the immiscible atoms seeking to avoid close contact in the solid-state. As proof of principle, we present the discovery, crystal growth, and basic physical properties of La$_4$Co$_4$\textit{X} (\textit{X} = Pb, Bi, Sb), a new family of intermetallic compounds based on the antagonistic pairs Co-Pb and Co-Bi. La$_4$Co$_4$\textit{X} adopts a new orthorhombic crystal structure (space group \textit{Pbam}) containing quasi-2D Co slabs and La-\textit{X} polyhedra that stack in an alternating manner along the \textit{a}-axis. Consistent with our proposal, the La atoms separate the Co and \textit{X} substructures, ensuring there are no direct contacts between the members of the immiscible (antagonistic) pair. Within the Co slabs, the atoms occupy the vertices of corner sharing tetrahedra and triangles, and this bonding motif produces flat electronic bands near the Fermi level that favor magnetism. The Co is moment bearing in each La$_4$Co$_4$\textit{X} compound studied, and we show that whereas La$_4$Co$_4$Pb behaves as a three dimensional antiferromagnet with \textit{T}$_{\text{N}}$ = 220 K, La$_4$Co$_4$Bi and La$_4$Co$_4$Sb have behavior consistent with low dimensional magnetic coupling and ordering, with \textit{T}$_{\text{N}}$ = 153 K and 143 K respectively. In addition to the Pb, Bi, and Sb based La$_4$Co$_4$\textit{X} compounds, we also were likely able to produce an analogous La$_4$Co$_4$Sn in polycrystalline form, although we were unable to isolate single crystals. We anticipate that identifying and using mutually compatible third elements together with an antagonistic pair represents a novel and generalizable design principle for discovering new materials and new structure types containing low-dimensional substructures. 


\end{abstract}

\end{center}

\vspace{7mm}

\end{@twocolumnfalse}
]

\section{Introduction}

\begin{figure}[!t]
    \centering
    \includegraphics[width=\linewidth]{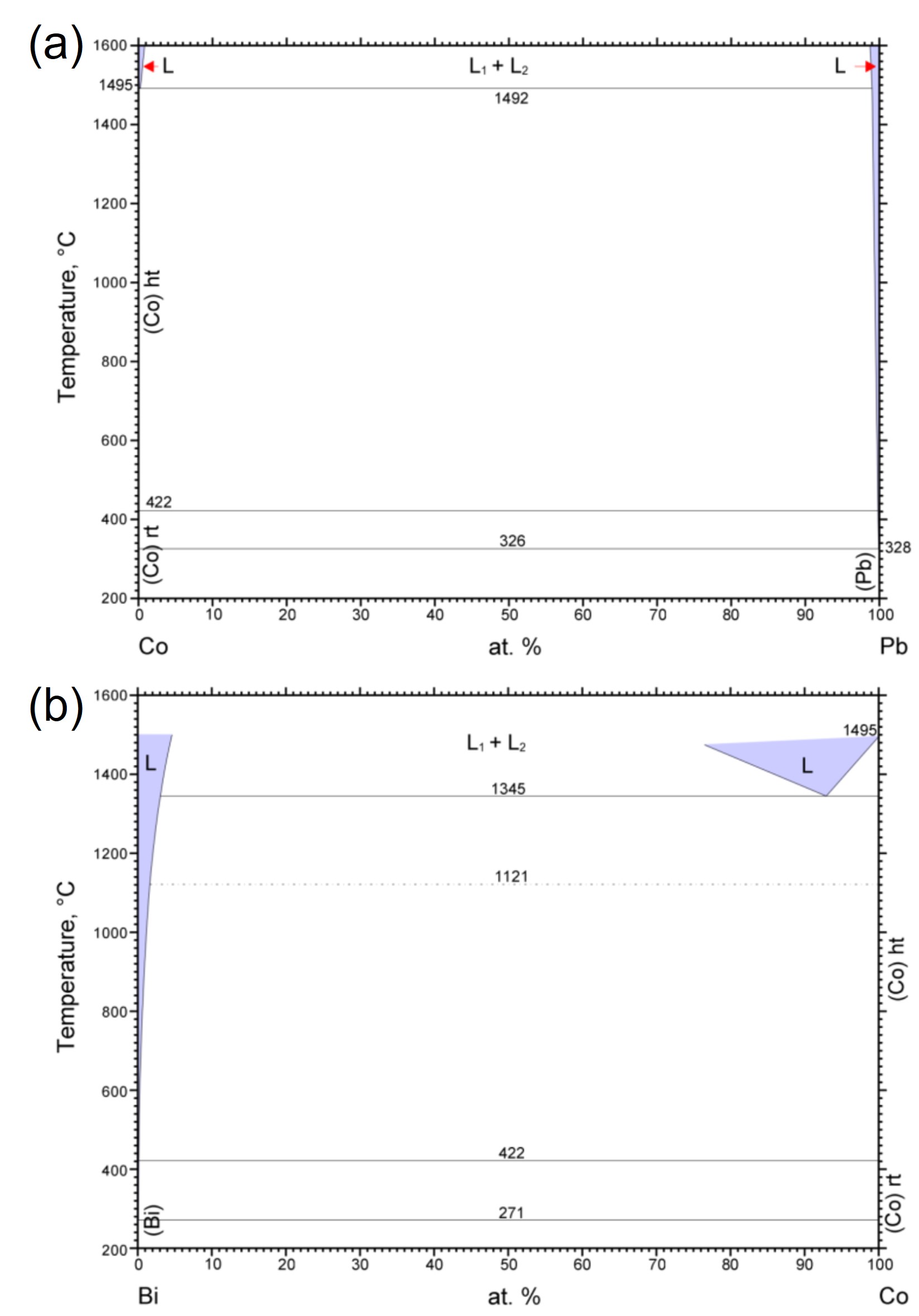}
    \caption[]{(a) Binary phase diagram for Co and Pb (ASM Diagram $\#$900732,\autocite{ASM_Co-Pb}). (b) Binary phase diagram for Co and Bi (ASM Diagram $\#$900429,\autocite{ASM_Bi-Co}). The diagrams demonstrate the extreme immiscibility of Co with Pb and Bi}
    \label{PhaseDiagram}
\end{figure}

A foundational aspiration of solid state chemistry is to develop general design principles for producing materials with targeted structural motifs and physical properties. Here, we demonstrate how strongly immiscible pairs of elements can be leveraged to synthesize ternary compounds with reduced or quasi-reduced dimensional motifs such as 2D sheets, 1D chains, or 0D clusters intrinsically built into their crystal structures. To introduce the basic idea, consider the two binary phase diagrams for Co-Pb and Co-Bi displayed in Figures \ref{PhaseDiagram}a and \ref{PhaseDiagram}b. The diagrams show that Co is extraordinarily immiscible with Pb and Bi at temperatures up to at least 1400°C, and extrapolating to higher temperatures, remain immiscible as liquids even at thousands of degrees. We use the term "antagonistic pairs" to refer to such pairs of elements with similarly extreme levels of immiscibility,\autocite{canfield2019new} of which other examples include, but are not limited to, Ba-Fe and La-Cr.

\begin{figure*}[!t]
    \centering
    \includegraphics[width=\linewidth]{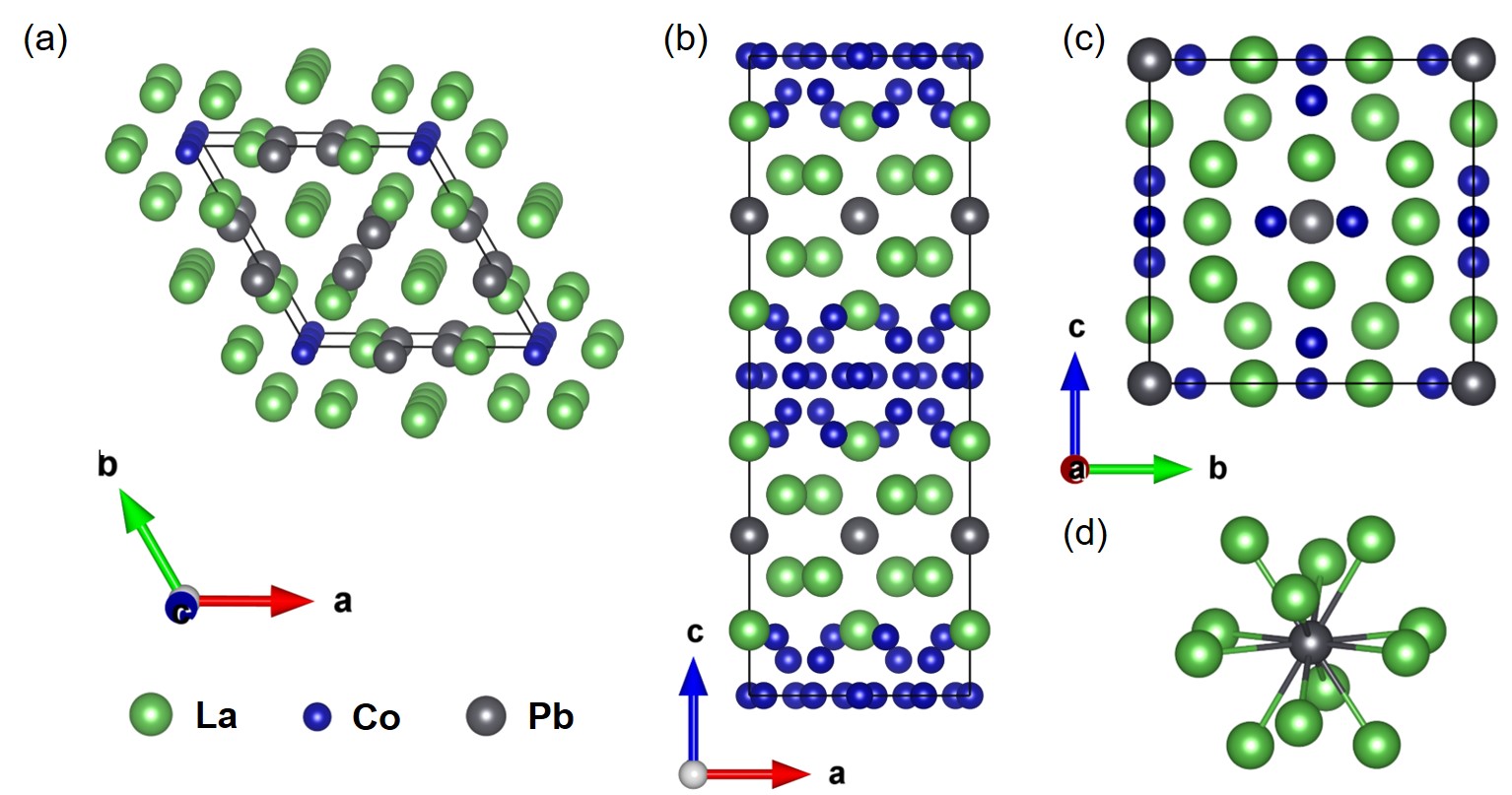}
    \caption[]{Crystal structures of the previously reported ternary compounds containing La, Co, and Pb. (a) La$_5$CoPb$_3$, La$_6$Co$_{13}$Pb, and (c) La$_{12}$Co$_{6}$Pb. (d) shows the local coordination environment of the Pb atoms in La$_{12}$Co$_{6}$Pb (pannel (c)), emphasizing the complete encapsulation of Pb by La with no Pb-Co contacts.}
    \label{La_Co_Pb_ternaries}
\end{figure*}

The heart of this work is the question: for any given immiscible pair of elements, what third element can be introduced to produce stable ternary compounds? Remarkably, a survey of crystallographic databases reveals that for many ternary compounds that contain an antagonistic pair, the strong immiscibility of the antagonistic elements often remains preserved in the crystal structure, with the third element separating the immiscible elements into distinct, quasi low dimensional substructures.\autocite{canfield2019new} As an example, the pair Co-Pb forms ternary compounds with most rare earth elements, and Figure \ref{La_Co_Pb_ternaries} illustrate the crystal structures for La$_5$CoPb$_3$, La$_6$Co$_{13}$Pb, and La$_{12}$Co$_{6}$Pb, which span the known structure types found in the broader \textit{R}-Co-Pb (\textit{R} = rare earth atom) phase space.\autocite{guloy1994exploration,weitzer1993magnetism,gulay2000crystal} The crystal structures each clearly reflect the immiscibility of the Co and Pb atoms. La$_5$CoPb$_3$ contains quasi-1D chains of Co atoms that extend down the \textit{c}-axis and are separated from the Pb by sheaths of La. The structure of La$_6$Co$_{13}$Pb can be envisioned as 2D slabs of Co and sheets of Pb that stack along the \textit{c}-axis and are separated by layers of La atoms. Finally, in La$_{12}$Co$_{6}$Pb, single, "zero-dimensional," Pb atoms are fully encapsulated by La such that there are no nearest neighbor Pb-Co contacts. 

The key insight from Figure \ref{La_Co_Pb_ternaries} is that in each structure, the addition of the mutually compatible third element, La, to the immiscibile pair, Co-Pb, produces ternary compounds in which the La atoms segregate the Co and the Pb, giving materials in which at least one of the atoms occupies a quasi-reduced dimensional substructure like the 1-D Co chains in La$_5$CoPb$_3$, 2-D Co slabs in La$_6$Co$_{13}$Pb, or 0-D Pb atoms in La$_{12}$Co$_{6}$Pb. Other examples based on different antagonistic pairs, such as \textbf{LaCr}Ge$_3$ and the \textbf{\textit{AE}Fe$_2$}As$_2$ (\textit{AE} = alkali earth) superconductors, display similar structural segregation and quasi-low dimensional motifs in their crystal structures (the immiscible pairs are in bold). Importantly, in systems containing magnetic, or potentially moment bearing atoms, such as 3\textit{d} transition metals, the reduced dimensionality may lead to enhanced fluctuations and emergent, correlated phases with exotic physical properties such as the unconventional superconductivity in the iron-arsenides\autocite{canfield2010feas,stewart2011superconductivity} or fragile, itinerant, magnetism in LaCeGe$_3$.\autocite{PhysRevLett.117.037207,kaluarachchi2017tricritical,PhysRevB.103.075111} 

We propose here that searching for ternary compounds based upon a strongly immiscible pair of elements constitutes a very general design principle for identifying and discovering novel materials with low dimensional structural motifs favorable for exhibiting emergent quantum properties. As a proof of principle, we outline the discovery of a new family of ternary compounds La$_4$Co$_4$\textit{X} (\textit{X} = Pb, Bi, Sb) that are based on the antagonistic pairs Co-Pb and Co-Bi. The La$_4$Co$_4$\textit{X} adopt a new structure type with an orthorhombic \textit{Pbam} arrangement. Like the compounds discussed above, the La$_4$Co$_4$\textit{X} structure reflects the strong Co-\textit{X} immisicibility, and consists of staggered, quasi-2D slabs of Co separated by layers of La-encapsulated \textit{X} atoms. The Co slabs can be visualized as chains of corner sharing tetrahedra that run down the \textit{c}-axis and are linked along the \textit{b}-direction by corner sharing triangles. The Co is moment bearing in each compound, with La$_4$Co$_4$Pb showing behavior typical of a three dimensional antiferromagnet with \textit{T}$_{\text{N}}$ = 220 K, whereas La$_4$Co$_4$Bi and La$_4$Co$_4$Sb 
have temperature dependent magnetization characteristic of low dimensional coupling and ordering, with \textit{T}$_{\text{N}}$ = 153 K and 143 K respectively. This work demonstrates that the extreme immiscibility of antagonistic element pairs can be overcome and exploited using a mututally compatible third element to produce new quantum materials and new structure types with built-in quasi-reduced dimensionality.


\section{Experimental Details}

\subsection{Crystal Growth} 

\begin{table*}[htbp]
  \centering
  \caption{Single crystal data and structural refinement information for La$_4$Co$_4$\textit{X} (\textit{X} = Pb, Bi, Sb).}
    \resizebox{\linewidth}{!}{\begin{tabular}{llll}
    \toprule
    Chemical formula & La$_4$Co$_4$Pb & La$_4$Co$_4$Bi & La$_4$Co$_4$Sb \\
    Formula weight (g/mol) & 998.55 & 1000.34 & 913.11 \\
    Temperature & 295.3(7) & 294.7(3) & 295.32(10) \\
    Wavelength (\AA, Ag K$\alpha$) & 0.56087 & 0.56087 & 0.56087 \\
    Crystal system & orthorhombic & orthorhombic & orthorhombic \\
    Space group & \textit{Pbam} (\# 55) & \textit{Pbam} (\# 55) & \textit{Pbam} (\# 55) \\
    \multirow{3}[0]{*}{Unit cell dimensions} & \textit{a} = 21.9909(3) \AA, $\alpha$ = 90° & \textit{a} = 22.1952(10) \AA, $\alpha$ = 90° & \textit{a} = 22.0031(4) \AA, $\alpha$ = 90° \\
        & \textit{b} = 8.28370(10) \AA, $\beta$ = 90° & \textit{b} = 8.3236(3) \AA, $\beta$ = 90° & \textit{b} = 8.2780(2) \AA, $\beta$ = 90° \\
        & \textit{c} = 4.74940(10) \AA, $\gamma$ = 90° & \textit{c} = 4.7224(2) \AA, $\gamma$ = 90° & \textit{c} = 4.71320(10) \AA, $\gamma$ = 90° \\
    Volume (cm$^3$) & 865.18(2) & 872.43(6) & 858.47(3) \\
    Z   & 4   & 4   & 4 \\
    Calculated density (g/cm$^3$) & 7.666 & 7.616 & 7.065 \\
    Absorption coefficient (mm$^{-1}$) & 24.366 & 24.868 & 15.721 \\
    Absorption Correction & Face indexed & multi-scan & Face indexed \\
    F(000) & 1672 & 1676 & 1548 \\
    Crystal size (mm) & 0.26 x 0.19 x 0.08 & 0.2 x 0.17 x 0.13 & 0.16 x 0.13 x 0.1 \\
    theta range for data collection & 2.4020-24.6820 & 2.4-28.9270 & 2.423-29.2900 \\
    Index ranges (min/max, h, k, l) & [-32/30, -11/11, -6/6] & [-35/34, -13/13, -7/8] & [-36/35, -13/13, -8,7] \\
    Reflections collected & 19831 & 30382 & 34091 \\
    Independent reflections & 1454 ($R_{\text{int}}$ = 0.0708) & 2411 ($R_{\text{int}}$ = 0.0625) & 2366 ($R_{\text{int}}$ = 0.0528) \\
    Completeness & 99.9 & 99.8 & 99.8 \\
    Refinement method & Full-matrix least-squares on F$^2$ & Full-matrix least-squares on F$^2$ & Full-matrix least-squares on F$^2$ \\
    Data / restrains / parameters & 2146 / 0 / 55 & 2411 / 0 / 55 & 2366 / 0 / 55 \\
    GOF & 1.286 & 1.232 & 1.253 \\
    Final \textit{R} indices [I $>$ 2$\sigma$(I)] & $R_{\text{obs}}$ = 0.0272, $wR_{\text{obs}}$ = 0.0652 & $R_{\text{obs}}$ = 0.0233, $wR_{\text{obs}}$ = 0.0429 & $R_{\text{obs}}$ = 0.0198, $wR_{\text{obs}}$ = 0.0385 \\
    \textit{R} indices [all data] & $R_{\text{all}}$ = 0.0279, $wR_{\text{all}}$ = 0.0655 & $R_{\text{all}}$ = 0.0263, $wR_{\text{all}}$ = 0.0434 & $R_{\text{all}}$ = 0.0221, $wR_{\text{all}}$ = 0.0390 \\
    Extinction coefficient & 0.00413(19) & 0.00122(6) & 0.00433(12) \\
    Largest diff. peak and hole (e$^-$/\AA$^3$) & 2.907 and -3.117 & 1.834 and -1.683 & 1.266 and -1.484 \\
    CSD Number & 2333895 & 2333982 & 2333786 \\
    \bottomrule
    \end{tabular}}%
  \label{crystal}%
\end{table*}%

Whereas Co is strongly immiscible with Pb and Bi, all three elements, Co, Pb, and Bi, mix readily with La.\autocite{okamoto1990binary} The Co-La binary phase diagram shows a relatively deep eutectic near $\approx$ 70 $\%$ La, and compositions near the eutectic point allow for access to single phase liquid at temperatures below 750°C for $\approx$ 45-85 $\%$ La,\autocite{okamoto1990binary} and therefore are attractive solvents from which to grow La-Co-\textit{X} ternary compounds.

We grew the La$_4$Co$_4$\textit{X} single crystals by adding small quantities of \textit{X} = Pb, Bi, Sb to Co-La based melts as follows. Elemental La (Ames Laboratory, 99.9+ $\%$) Co pieces (American Elements, 99.99 $\%$), and Pb (Alfa Aesar, 99.99 $\%$), Bi (Alfa Aesar, 99.99 $\%$), or Sb (Alfa Aesar, 99.99 $\%$) were weighed in molar ratios of La$_{45}$Co$_{45}$Pb$_{10}$, La$_{51}$Co$_{46}$Bi$_{3}$, and La$_{43}$Co$_{53}$Sb$_{4}$ and placed in a Ta crucible set with home made Ta caps and filter.\autocite{canfield2001high,canfield2016use} The Ta crucibles were sealed under an Ar atmosphere using an arc melter, and then the crucibles were flame sealed under vacuum in fused silica ampoules. The ampoules were heated in a box furnace to 1150°C. After dwelling at 1150°C for 10 h, the furnace was cooled to 775°C over 100-200 h after which the samples were removed from the furnace and the excess flux was decanted in a centrifuge with metal cups and rotors.\autocite{canfield2019new} After cooling to room temperature, the tubes were opened to reveal a mixture of plate-like crystals and smaller, thin blade-like rods for samples containing Pb and Sb, whereas the Bi containing samples only yielded the blades. The inset to Figure \ref{MT_RT}a shows typical example of the blade-like crystals. We emphasize that the above compositions were those used to grow the specific crystals whose data is presented in the following sections. Other, similar, ratios of La, Co, and \textit{X} were also used in an attempt to increase the crystal size and/or phase purity. These other attempts generally produced qualitatively similar mixtures of plates and blades.

EDS and powder x-ray diffraction indicated the larger plates to be the known ternary La$_6$Co$_{13}$\textit{X} (\textit{X} = Pb or Sb)\autocite{weitzer1993magnetism} and the blade-like rods to be a new ternary compound with the chemical formula La$_4$Co$_4$\textit{X} (\textit{X} = Pb, Bi, or Sb). The crystals were found to be oxygen and/or moisture sensitive, and the bright, metallic luster of the as-grown samples darkened even after several minutes of air exposure. Therefore, the samples were stored in a nitrogen glovebox until needed for characterization. We also attempted to grow a \textit{X} = Sn member. Although we were unable to isolate suitable crystals for further characterization, our polycrystalline samples had clear evidence of a La$_4$Co$_4$Sn phase. More details on our attempts to grow La$_4$Co$_4$Sn are outlined in the supporting information. 

\subsection{Elemental analysis}

The composition of the blade-like samples was determined by Energy Dispersive Spectroscopy (EDS) quantitative chemical analysis using a ThermoFisher (FEI) Teneo Lovac FE-scanning electron microscope (SEM). The data was analyzed using an Oxford Instruments Aztec System with X-Max-80 detector, attached to the Teneo. The measurements were conducted with an acceleration voltage of 20 kV, current of 1.6 nA, working distance of 10 mm, and a take off angle of 35°. We analyzed three separate crystals from each batch, respectively corresponding to Pb, Bi, and Sb containing samples, and the composition of each crystal was measured at 4-8 different spots, revealing good homogeneity in each case. The standards used for reference are internal to the Oxford software. The EDS data for each sample is summarized in Tables S1-S3 in the supporting information.

\subsection{X-ray diffraction} 

Single crystal X-ray diffraction was performed using a Rigaku XtaLab Synergy-S diffractometer with Ag  radiation (0.56087 \AA\ ) operating at 65 kV and 0.67 mA. The samples were held in a nylon loop with vacuum grease, and the data was collected at room temperature. The total number of runs and images was based on the strategy calculation from the program CrysAlisPro (Rigaku OD, 2023). The data integration and reduction were also performed using CrysAlisPro, and a numerical absorption correction was applied based on Gaussian integration over a face-indexed crystal. For La$_4$Co$_4$Bi, the crystal was irregularly shaped and we were unable to index the faces, so we employed an empirical multi-scan absorption correction using spherical harmonics, implemented in the SCALE3 ABSPACK scaling algorithm. The structures were solved by intrinsic phasing using the SHELXT software package and were refined with SHELXL. The crystallographic data and refinement results are given Tables \ref{crystal}, and the atomic positions and thermal displacement parameters are listed in Tables S4-S9 in the supporting information.

Powder X-ray diffraction patterns were obtained using a Rigaku Miniflex-II instrument operating with Cu-\textit{K}$\alpha$ radiation with $\lambda$ = 1.5406 \AA\ (\textit{K}$\alpha$1) and 1.5443 \AA\ (\textit{K}$\alpha$2) at 30 kV and 15 mA. The samples were prepared by grinding a representative number of crystals (5-10) to a fine powder. The powder patterns were refined using the Rietveld method with GSAS-II software.\autocite{toby2013gsas} To address air sensitivity, we also measured the powders on a Rigaku Miniflex-I instrument (same Cu-radiation and wavelength) contained in a nitrogen filled glovebox. The patterns are shown in Figures S1 and S2 in the supplemental information and were not found to be significantly different when comparing measurements conducted in air or nitrogen environments.

\subsection{Physical property measurements} 

\begin{figure*}[!t]
    \centering
    \includegraphics[width=\linewidth]{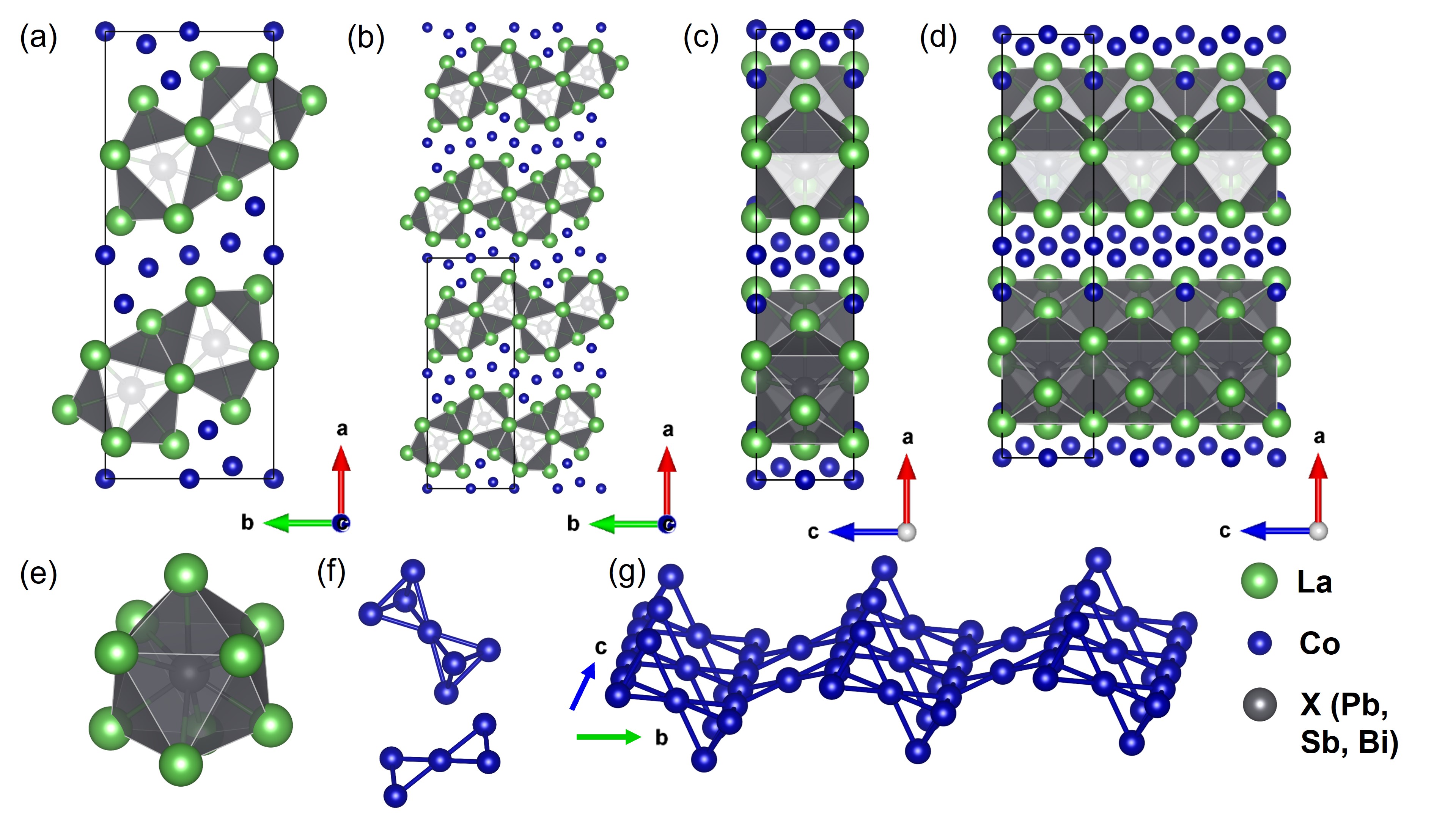}
    \caption[]{Crystal structure of La$_4$Co$_4$\textit{X} (\textit{X} = Pb, Bi, Sb). (a) Unit cell of La$_4$Co$_4$\textit{X} viewed down the \textit{c}-axis. (b) Extended view of the structure in the same orientation as (a) emphasizing the alternating stacks of Co-slabs and La-\textit{X} polyhedral layers along the \textit{a}-direction. (c) Unit cell viewed down the \textit{b}-axis. (d) Extended view of the orientation in (c) showing the face-sharing connectivity of the La-\textit{X} polyhedra along the \textit{b}-direction and the quasi-layered, distinct Co- and La$_9$\textit{X}-based structrual units. (e) Isolated coordination environment of \textit{X} surrounded by La to form tri-capped trigonal prisms. (f) Corner sharing tetrahedra and triangle sub-units that make up the Co slabs. (g) Connectivity of the two structural sub-units shown in (f) into extended Co-slabs that span the \textit{bc}-plane.}
    \label{structure}
\end{figure*}

Temperature dependent resistance measurements were performed between 4-300 K in a closed-cycle cryostat (Janis SHI-950). The AC resistance was measured using LakeShore AC resistance bridges (models 370 and 372), with a frequency of 17 Hz and a 3 mA excitation current. The temperature was measured by a calibrated Cernox 1030 sensor connected to a LakeShore 336 controller. As the crystals naturally grew with a blade-like morphology, we measured the resistances along the long-axis of the blade, which was determined with a Laue camera to be the \textit{c}-crystallographic axis. To prepare the samples, the surfaces of the crystals were lightly polished, and contacts were made by spot welding 25 $\mu$m thick annealed Pt wire onto the (100) faces of the crystals in standard four point geometry. After spot welding, a small amount of silver epoxy was painted onto the contacts to ensure good mechanical strength, and typical contact resistances were $\approx$ 1 $\Omega$.

Magnetization measurements were performed in a Quantum Design Magnetic Property Measurement System (MPMS-classic) SQUID magnetometer operating in the DC measurement mode. The measurements were performed with the field applied along the \textit{a}, \textit{b}, and \textit{c} crystallographic directions, and the samples were oriented with a Laue camera. For La$_4$Co$_4$Pb and La$_4$Co$_4$Bi, the samples were mounted between two straws for measurements in the \textit{H} $\parallel$ \textit{b} and \textit{H} $\parallel$ \textit{c} orientations. To measure with \textit{H} $\parallel$ \textit{a} (perpendicular to the thinnest direction of the blade-like crystals), the samples were glued to a Kel-F disc, which was then placed inside of a straw. Prior to measuring, the blank disc was first measured at the same temperatures and fields to use as a background subtraction. For La$_4$Co$_4$Sb, the sample was mounted to the disc to measure all orientations.

\subsection{Electronic structure calculations} 

Density functional theory (DFT) was used to calculate the electronic structure of La$_4$Co$_4$\textit{X} with \textit{X} = Pb, Bi, and Sb. The DFT calculations were performed by using the Perdew-Burke-Ernzerhof (PBE) exchange-correlation functional within the framework of GGA\autocite{PhysRevLett.77.3865} as implemented in the Vienna Ab initio Simulation Package (VASP).\autocite{kresse1996efficiency,PhysRevB.54.11169} A plane-wave cutoff energy was set to 520 eV, the experimental lattice parameters were used, and the positions of atoms were fully optimized with the force tolerance of 0.01 eV/\AA\. The accuracy of the electron self-consistent field is set to 10$^{-4}$ eV, and the Brillouin zone is sampled using a set of gamma-centered uniform 2 $\times$ 8 $\times$ 16 grids with tetrahedron method for the density of states calculations. 

\section{Results and Discussion}

\begin{figure*}[!t]
    \centering
    \includegraphics[width=\linewidth]{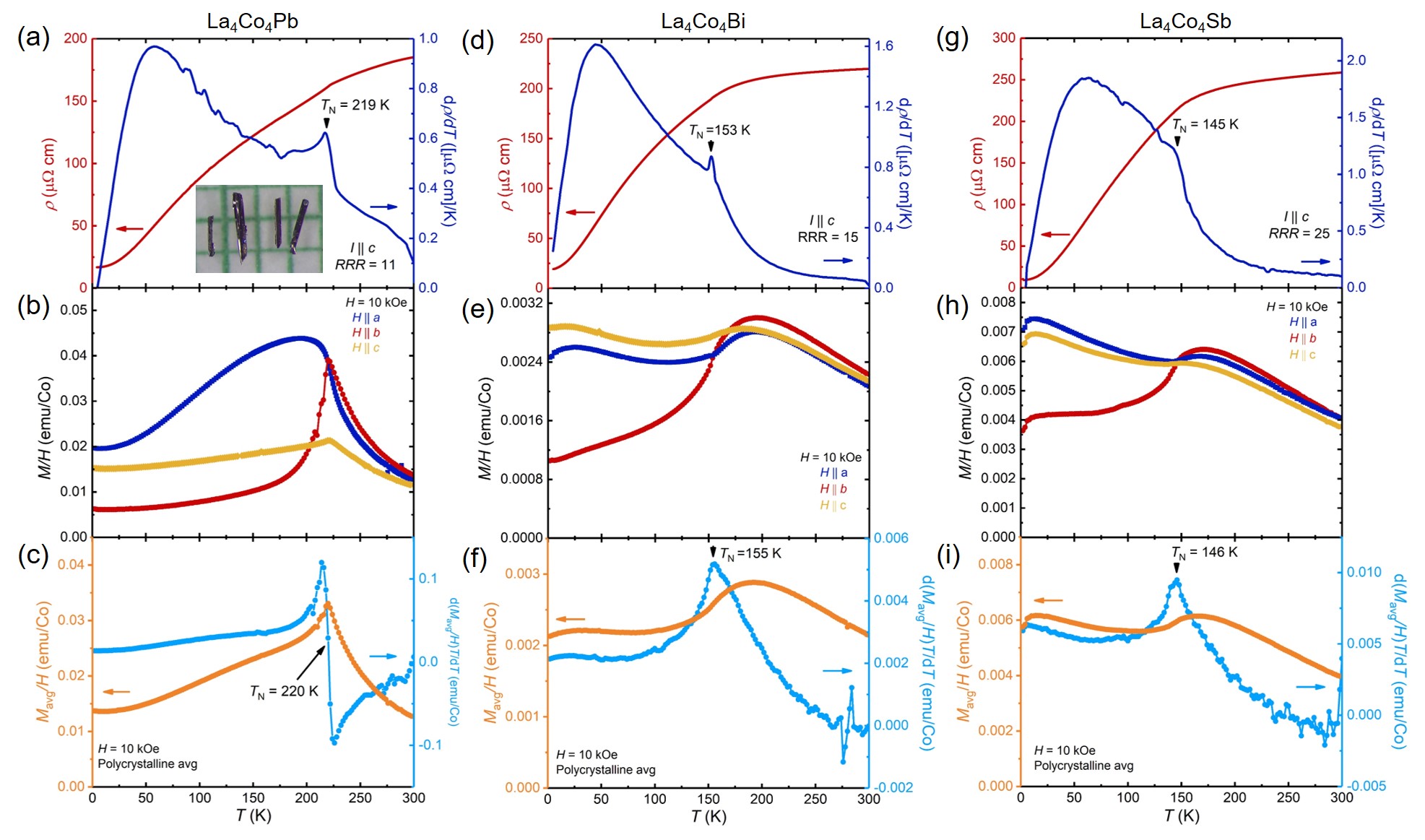}
    \caption[]{(a) Temperature dependent resistivity (left axis, in red) and its derivative d$\rho$/d\textit{T} (right axis, in blue), (b) Anisotropic temperature dependent magnetization (\textit{M}/\textit{H}), and (c) polycrystalline average of \textit{M/H} (left axis, orange) and d[(\textit{M}$_{avg}$/\textit{H})\textit{T}]/d\textit{T} (right axis, light blue) of the magnetization data in (b). (a)-(c) show the data for La$_4$Co$_4$Pb, (d-f) the data for La$_4$Co$_4$Bi, and (g-i) the data for La$_4$Co$_4$Sb. The inset in (a) shows a picture of typical La$_4$Co$_4$Pb crystals on a mm grid.}
    \label{MT_RT}
\end{figure*}

\begin{figure}[!t]
    \centering
    \includegraphics[width=\linewidth]{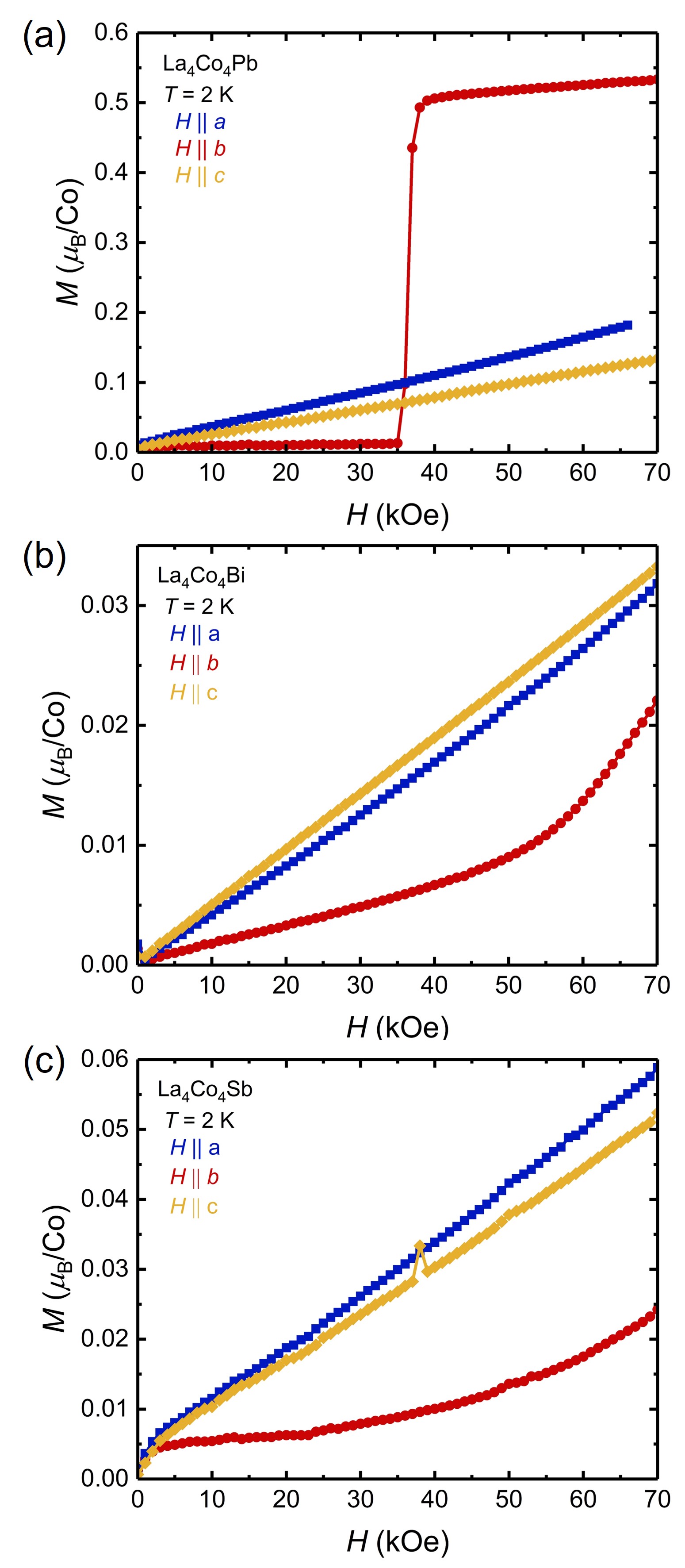}
    \caption[]{Field dependent magnetization isotherms measured at \textit{T} = 2 K for (a) La$_4$Co$_4$Pb, (b) La$_4$Co$_4$Bi, and (c) La$_4$Co$_4$Sb.}
    \label{MH_2K}
\end{figure}

\begin{figure*}[!t]
    \centering
    \includegraphics[width=\linewidth]{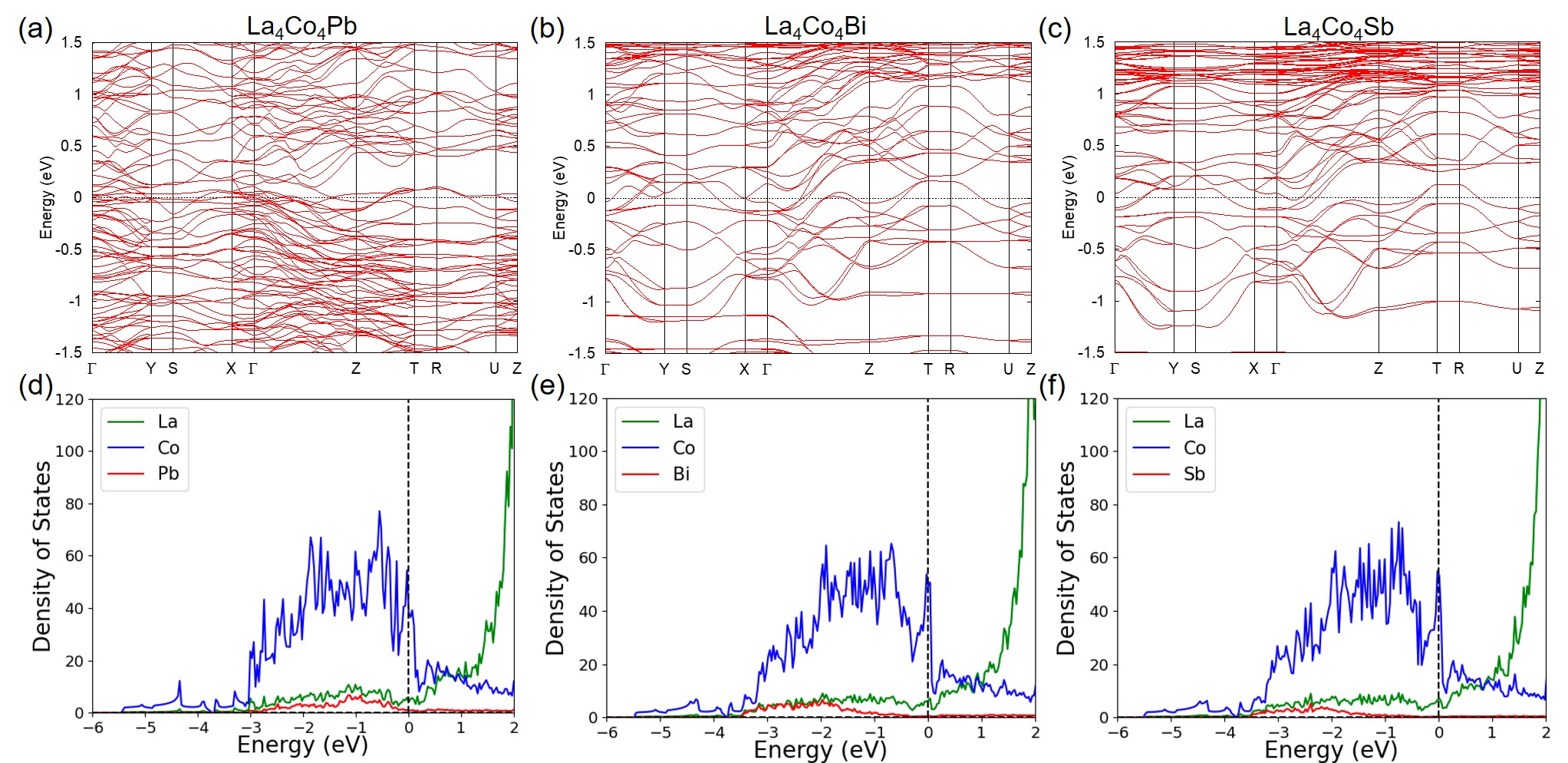}
    \caption[]{DFT calculated electronic band structures for (a) La$_4$Co$_4$Pb, (b) La$_4$Co$_4$Bi, and (c) La$_4$Co$_4$Sb. (d-f) show the calculated density of states for the corresponding band structures in (a-c).}
    \label{bands}
\end{figure*}

The La$_4$Co$_4$\textit{X} (\textit{X} = Pb, Bi, or Sb) compounds adopt a new orthorhombic structure type with space group \textit{Pbam}. The crystal structure is illustrated in Figure \ref{structure}, and information regarding the structural refinements is given in Table \ref{crystal}. The atomic positions and thermal displacement parameters are summarized in Tables S4-S9 in the supporting information. As shown in Figures \ref{structure}a and \ref{structure}b, La$_4$Co$_4$\textit{X} adopts a quasi-layered structure featuring striking segregation of the immiscible Co and \textit{X} atoms into two distinct structural sub-units, Co slabs and slabs formed from La$_9$\textit{X} polyhedra. The Co and La-\textit{X} slabs each span the \textit{bc}-plane and stack in an alternating manner along the \textit{a}-axis. Within the La-\textit{X} slabs, the \textit{X} atoms are nine-fold coordinated at the center of tricapped trigonal prisms (see Figure \ref{structure}e), and the polyhedra have face-sharing connectivity that extends linearly down the \textit{c} axis and is staggered along the \textit{b}-direction, as emphasized in Figures \ref{structure}c and \ref{structure}d. 

Figure \ref{structure}f shows that the Co atoms sit at the vertices of two structural sub-units within the greater Co-slabs: pairs of corner sharing tetrahedra and corner sharing triangles. The tetrahedra extend in chains down the \textit{c}-axis and are linked in the \textit{b}-direction by the corner sharing triangles to give the network shown in Figure \ref{structure}g. An alternative way to view the Co-slabs is as a corrugated Kagome lattice that is bent in a zig-zag fashion along the \textit{b}-axis with additional Co atoms (Co3, see atomic positions listed in Tables S4, S6, and S8 supporting information) sitting above and below the corrugated kagome layers.

The existence of stable ternary compounds containing the pairs Co-Pb and Co-Bi demonstrates that the extreme immiscibility of Co with Pb and Bi may be overcome by using a third element that is mutually compatible with the antagonistic elements. In this case, Co and Pb/Bi are all soluble in excess La at temperatures above 800° C.\autocite{okamoto1990binary} Within the La$_4$Co$_4$\textit{X} crystal structure, the La clearly acts to spatially segregate the Co and \textit{X}, as the complete encapsulation of \textit{X} into the La$_9$\textit{X} polyhedra ensures the \textit{X} atoms are chemically isolated from the Co slabs. As the La$_4$Co$_4$\textit{X} family are new materials that adopt a new structure type, this work serves as a proof-of-principle that highly immiscible pairs of atoms can be leveraged to form new materials with quasi-reduced dimensional motifs intrinsically built into their crystal structures, which we suggest naturally form as immiscible atoms continue to avoid close contact in the solid-state. 

In La$_4$Co$_4$\textit{X}, the most interesting structural motifs are the quasi-2D slabs of Co the span the \textit{bc}-plane. As described above, these slabs are composed of corner sharing tetrahedra that form chains along the \textit{c}-axis and that are linked by pairs of corner sharing triangles along the \textit{b}-direction. Such networks of corner sharing tetrahdra or triangles form the basis of pyrochlore and Kagome lattices, both of which are currently attracting immense interest as potential platforms for studying many-body topological physics.\autocite{cualuguaru2022general,regnault2022catalogue,PhysRevA.103.L031301,wang2018large,han2021evidence,kang2020topological,ye2018massive,ye2018massive,ye2021flat,neves2023crystal} In simplified models for metals with pyrochlore and Kagome lattices, frustrated hopping from adjacent sites leads to quenching of the electron kinetic energy and electronic localization. This produces flat electronic bands in momentum space and increases the relative strength of the Coulomb interaction between electrons, potentially providing a driver for correlated states including itinerant magnetism,\autocite{teng2023magnetism} charge density waves,\autocite{teng2023magnetism,PhysRevMaterials.3.094407,jiang2021unconventional,PhysRevX.11.031026} and unconventional superconductivity.\autocite{balents2020superconductivity,PhysRevLett.126.027002,PhysRevMaterials.5.034801,zhao2021cascade,guguchia2023tunable}

To determine if the La$_4$Co$_4$\textit{X} family shows interesting physical properties, we measured the temperature and field dependence of the resistivity and magnetization of each. The data is presented in Figure \ref{MT_RT}. The top row, Figures \ref{MT_RT}a, \ref{MT_RT}d, and \ref{MT_RT}g show the resistivity ($\rho$, left axis, in red) and its derivative d$\rho$/d\textit{T} (right axis, in blue) of each La$_4$Co$_4$\textit{X} compound. The La$_4$Co$_4$\textit{X} materials all show metallic behavior where the resistivity decreases with cooling between 4-300 K. The residual resistivity ratios, $\rho$(300 K)/$\rho$(4 K), are reasonably high, spanning 10-25, indicating good crystal quality. At high temperatures, the resistivity of each material has a modest slope, but each shows a kink on cooling that suggests a loss of spin disorder scattering as the samples enter a magnetically ordered state. The peaks in the resistance derivatives were used to assign the transition temperatures of \textit{T}$_N$ = 219 K for La$_4$Co$_4$Pb, 153 K for La$_4$Co$_4$Bi, and 145 K for La$_4$Co$_4$Sb,\autocite{fisher1968resistive} and we find that the d$\rho$/d\textit{T} peak is substantially stronger in La$_4$Co$_4$Pb than either La$_4$Co$_4$Bi or La$_4$Co$_4$Sb. We will return to this point when discussing the magnetic data.

Figures \ref{MT_RT}b, \ref{MT_RT}e, and \ref{MT_RT}h display the anisotropic temperature dependent magnetization, shown as \textit{M/H}. The magnetic data confirm that Co is moment bearing and orders in each La$_4$Co$_4$\textit{X} compound. Like in the resistivity data, the transition is most apparent (sharpest) in La$_4$Co$_4$Pb. As shown in Figure \ref{MT_RT}b, the \textit{H} $\parallel$ \textit{b} and \textit{H} $\parallel$ \textit{c} magnetization of La$_4$Co$_4$Pb first increases on cooling and reaches a peak near 220 K, characteristic of antiferromagnetic order in which the moments are aligned along the \textit{b}-axis. Above the transition, \textit{M/H} increases in a Curie-Weiss like manner, with substantial anisotropy between different orientations. Curie-Weiss fits, shown in Figure S6 in the supporting information, to the the \textit{M}/\textit{H} data suggest an effective moment of $\mu_{\text{eff}}$ = 3.5(1) $\mu_{\text{B}}$/Co and Weiss temperatures $\theta$ = 166(3) K, which is slightly smaller than, but in reasonable agreement with the 3.87 $\mu_{\text{B}}$ expected for Co$^{2+}$ moments.


In the Bi and Sb containing compounds, Figures \ref{MT_RT}e and \ref{MT_RT}h show that \textit{M/H} first increases with cooling but with a non Curie-Weiss like, essentially linear, temperature dependence. Likewise, instead of a sharp peak like that observed in La$_4$Co$_4$Pb, the magnetization of La$_4$Co$_4$Bi and La$_4$Co$_4$Sb reaches a very broad maximum centered respectively at 195 K and 170 K in each material. Below the maximum, \textit{M/H} decreases rapidly when \textit{H} $\parallel$ \textit{b}, but only subtly in the \textit{H} $\parallel$ \textit{a} and \textit{H} $\parallel$ \textit{c} orientations before settling into a relatively weak, slightly positive temperature dependence. The fact that the \textit{M/H} is smallest for the \textit{H} $\parallel$ \textit{b} orientation in each sample suggests that the ordered moments may be aligned primarily along the \textit{b}-axis within the antifferomagnetic state. We also note that the small downturn in \textit{M/H} observed for La$_4$Co$_4$Sb is an artifact from subtracting the disc (see experimental details), and is likely apparent owing to the small, $\approx$ 0.5 mg, sample mass and resulting weak signal. 

To estimate the Neel temperatures from the magnetic data, we calculated the polycrystalline average of the anisotropic data (shown in Figures \ref{MT_RT}c, \ref{MT_RT}f, and \ref{MT_RT}i, left axis in orange), where \textit{M}$_{\text{avg}}$/\textit{H} = 1/3(\textit{M}$_{\text{a}}$ + \textit{M}$_{\text{b}}$ + \textit{M}$_{\text{c}}$)/\textit{H}. The Neel temperatures were then determined using the derivatives d[(\textit{M$_{\text{avg}}$/H)T}]/d\textit{T},\autocite{fisher1962relation} which are presented in Figures \ref{MT_RT}c, \ref{MT_RT}f, and \ref{MT_RT}i (right axis, in light blue). The inferred \textit{T}$_{\text{N}}$ agree well with the transition temperatures determined from the resistivity derivatives, with \textit{T}$_N$ = 220 K, 155 K, and 146 for respective \textit{X} = Pb, Bi and Sb samples. Like in the d$\rho$/d\textit{T} data, the d[(\textit{M$_{\text{avg}}$/H)T}]/d\textit{T} behaves differently at \textit{T}$_N$ for La$_4$Co$_4$Pb, in which d[(\textit{M$_{\text{avg}}$/H)T}]/d\textit{T} has a sharp step/discontinuity corresponding to the maximum in \textit{M/H}. Instead of a discontinuity, the d[(\textit{M$_{\text{avg}}$/H)T}]/d\textit{T} curves for La$_4$Co$_4$Bi and La$_4$Co$_4$Sb have peaks at the respective temperatures that \textit{M$_{\text{avg}}$/H} reaches its greatest decreasing slope, approximately 20 K below the respective broad \textit{M/H} maxima. 

Figure \ref{MH_2K} shows the field dependent magnetization collected at a base temperature of 2 K for each compound. Corresponding data collected at 300 K is shown in Figure S5 in the supporting information and is consistent with paramagnetic behavior of the bulk, La$_4$Co$_4$\textit{X}. At 2 K, the low-field magnetization is smallest for all samples when the field is applied along the \textit{b}-axis. In the \textit{H} $\parallel$ \textit{a} and \textit{H} $\parallel$ \textit{c} orientations, the magnetization isotherms for each material have a near linear field dependence up to at least 70 kOe, and the samples all show metamagnetic transitions when \textit{H} $\parallel$ \textit{b}. Like the \textit{M/H} data, the isotherms shown in Figure \ref{MH_2K} again demonstrate that La$_4$Co$_4$Pb behaves qualitatively different than \textit{X} = Bi and Sb samples, with a much sharper feature in the \textit{M}(\textit{H}) isotherm. In La$_4$Co$_4$Pb, the \textit{b}-axis magnetization is extremely flat, and essentially independent of field up to 35 kOe, upon which it undergoes a step-like transition to a high-field state in which the magnetization is again nearly independent of field, with a saturated value of 0.53 $\mu_{\text{B}}$/Co. In La$_4$Co$_4$Bi and La$_4$Co$_4$Sb, the \textit{H} $\parallel$ \textit{b} isotherms also show a field induced transition beginning near 55-60 kOe that is incomplete at our maximum field of 70 kOe. However, the transition in the \textit{X} = Bi and Sb compounds appears substantially broadened compared to that observed in La$_4$Co$_4$Pb. Finally, we note that for La$_4$Co$_4$Sb, the small upturn in each isotherm below 5 kOe indicates this sample contains a small quantity of a ferromagnetic impurity (likely La$_6$Co$_{13}$Sb with \textit{T}$_C$ = 490 K\autocite{weitzer1993magnetism}), and a similar feature is observed in data collected at 300 K for this sample (see Figure S5 supporting information).


Taken together, La$_4$Co$_4$Pb exhibits distinct magnetic behavior from that of La$_4$Co$_4$Bi and La$_4$Co$_4$Sb. The magnetization of La$_4$Co$_4$Pb, with Curie-Weiss like \textit{M/H} in the high temperature paramagnetic regime and a well defined maximum at the Neel temperature, is typical of an intermetallic antiferromagnet with three dimensional coupling and ordering. La$_4$Co$_4$Bi and La$_4$Co$_4$Sb differ on both accounts, showing a broad maximum in \textit{M/H} instead of a sharp peak and non Curie-Weiss like temperature dependence in the paramagnetic regime. Furthermore, at the temperatures and fields applied here, the overall magnitude of \textit{M} for \textit{X} = Bi and Sb members is substantially smaller than that measured for La$_4$Co$_4$Pb. All of these characteristics are similar to the behavior of low dimensional systems, such as chains or sheets of coupled spins.\autocite{vasiliev2018milestones,de2001experiments}


More generally, the La$_4$Co$_4$\textit{X} family contains at least three new Co-based intermetallic compounds in which the Co is moment bearing. Compared to the rare-earth atoms, which provide robust magnetism associated with their well-localized 4\textit{f} orbitals, transition metals are considerably less likely to possess a magnetic moment in intermetallic compounds. Instead, the stronger hybridization between \textit{d}-orbitals (compared to 4\textit{f}) and the \textit{s}- and/or \textit{p}-like conduction electrons often leads to electronic delocalization and Pauli-paramagnetic behavior. In further contrast to the local moment physics found in rare-earth based materials, the magnetism of metals containing 3\textit{d} elements is often itinerant. In these cases, the on-sight Coulomb interaction and density of states at the Fermi level (\textit{E}$_{\text{F}}$) must be sufficient to spin-slit the bands near \textit{E}$_{\text{F}}$, leading to a net magnetic moment associated with the delocalized conduction electrons.\autocite{samolyuk2008relation,Santiago_2017}

In La$_4$Co$_4$\textit{X} it is possible that the bonding within the quasi-2D Co network, consisting of chains of corner sharing tetradedra linked by corner sharing triangles, produces flat electronic bands near the Fermi level that favor itinerant magnetism, similar to what is observed in some metals with an ideal kagome or pyrochlore sublattice.\autocite{kang2020topological,kang2020dirac,teng2023magnetism,ye2021flat,neves2023crystal} To explore this possibility, we calculated the nonmagnetic electronic band structure and density of states (DOS) of each La$_4$Co$_4$\textit{X} material, which are shown in Figure \ref{bands}. The calculations reveal each compound has relatively flat electronic bands very close to the Fermi level. Figures \ref{bands}d-\ref{bands}f show that the flat bands near \textit{E}$_{\text{F}}$ are derived from Co atoms. Whereas the calculations suggest all three La$_4$Co$_4$\textit{X} compounds have an enhanced DOS near \textit{E}$_{\text{F}}$, both the band structures and DOS features are different between La$_4$Co$_4$Pb and La$_4$Co$_4$Pn (Pn = Bi, Sb). The band structures of the two pnictogen containing compounds are very similar, and feature a very narrow spike in the density of states at \textit{E}$_{\text{F}}$. The bands of La$_4$Co$_4$Pb qualitatively differ, and the local maxima in the density of states is broader. These differences likely reflect the greater chemical similarity between La$_4$Co$_4$Bi and La$_4$Co$_4$Sb compared to La$_4$Co$_4$Pb, and are at least consistent with the experimental transport and magnentic data, in which behavior of La$_4$Co$_4$Pb is distinct. Finally, we emphasize that whereas our calculations indeed show flat, Co-based, bands in each La$_4$Co$_4$\textit{X} material, at present, we are unable to distinguish between the cases of trivial narrow bands, coming from poor orbital overlap or topological flat bands arising from destructive, out of phase hopping within the Co slabs. We will pursue this distinction in future work, with an eye towards understanding the conditions under which frustrated lattices which deviate from the perfect kagome or pyrochlore arrangement may still produce topological flat bands. In either case, the narrow bands and local maxima in the density of states near \textit{E}$_{\text{F}}$ are both consistent with the expectations for an intermetallic magnet based on 3\textit{d} elements.

Overall, our measurements show that La$_4$Co$_4$\textit{X} is a new family of Co based itinerant magnets in which quasi-2D slabs of Co produce magnetic behavior that is likely sensitive to the band filling. It will be worthwhile to pursue deliberate substitution work, i.e. La$_4$Co$_4$Pb$_{1-x}$Bi$_x$ to carefully study how the magnetism changes as a function of the electron count. Furthermore, we intend to also search for isostructural analogues containing moment bearing rare earth atoms to study the more complex magnetic phases likely accessible in such systems.

\section{Summary and Conclusions}

We highlighted how immiscibility can be exploited to target ternary compounds with low dimensional substructures, such as sheets, chains, or clusters intrinsically built into their crystal structures. As a demonstration, we present the discovery of a new family of intermetallic compounds, La$_4$Co$_4$\textit{X} (\textit{X} = Pb, Bi, Sb), that are based on the immiscible pairs Co-Pb and Co-Bi. La$_4$Co$_4$\textit{X} adopts a new orthorhombic structure with space group \textit{Pbam}. The La atoms divide the crystal structure into spatially separated Co and \textit{X} sub-units, where the Co atoms form quasi-2D slabs that span the \textit{bc}-plane and stack along the \textit{a}-axis, where they alternate with slabs of face sharing La$_9$\textit{X} polyhedra. Co is moment bearing in each material, and La$_4$Co$_4$Pb behaves as a three dimensional antiferromagnet with \textit{T}$_N$ = 220 K, whereas La$_4$Co$_4$Bi and La$_4$Co$_4$Sb both show behavior more consistent with low dimensional magnetic coupling and order marked by a broad maximum in the magnetization above their respective transitions at 153 K and 143 K. We also have been able to tentatively identify isostructural La$_4$Co$_4$Sn in polycrystalline form but have not yet isolated single phase samples. This work opens the door to a generalizable strategy of exploiting strongly immiscible pairs to target materials with low dimensional structural motifs. By further studying the chemistry dictating what third elements can be introduced to any given antagonistic pair to produce stable ternary compounds, we believe this concept represents a new design principle for the discovery of new materials and new structure types favorable for exhibiting emergent quantum behavior.

\section{Acknowledgements}

Work at Ames National Laboratory was supported by the U.S. Department of Energy, Office of Science, Basic Energy Sciences, Materials Sciences and Engineering Division. Ames National Laboratory is operated for the U.S. Department of Energy by Iowa State University under Contract No. DE-AC02-07CH11358. All electron microscopy and related work were performed using instruments in the Sensitive Instrument Facility at Ames National Laboratory. M.D. was supported by the U.S. Department of Energy, Office of Science, Office of Workforce Development for Teachers and Scientists (WDTS) under the Science Undergraduate Laboratory Internships (SULI) program. We especially thank Prof. Kirill Kovnir for valuable advice and discussions regarding the structure solutions and refinements.

\vskip 0.25cm
\noindent
*corresponding authors' email: slade@ameslab.gov, canfield@ameslab.gov

\vskip 0.25cm
\noindent
\textbf{\textit{Conflicts of Interest}}

The authors have no conflicts of interest to declare.

\printbibliography

\end{singlespace}

\end{document}


\begin{singlespace}

\title{\textbf{Supporting Information} \\
\vskip 0.5cm
La$_4$Co$_4$X (X = Pb, Bi, Sb): a demonstration of antagonistic pairs as a route to quasi-low dimensional ternary compounds}
\author{Tyler J. Slade$^{1*}$, Nao Furukawa$^{1,2}$, Matthew Dygert$^{1,3}$, Siham Mohamed,$^{1,4}$ Atreyee Das$^{1,2}$, \\ Weiyi Xia$^{1,2}$, Cai-Zhuang Wang$^{1,2}$, Sergey L. Bud’ko$^{1,2}$, Paul C. Canfield$^{1,2*}$}
\date{}

\maketitle

\begin{center} 
    
\textit{$^{1}$Ames National Laboratory, US DOE, Iowa State University, Ames, Iowa 50011, USA} \\  
\textit{$^{2}$Department of Physics and Astronomy, Iowa State University, Ames, Iowa 50011, USA} \\
\textit{$^{3}$Department of Physics and Astronomy, University of Missouri, Columbia, Missouri 65211, USA}\\
\textit{$^{4}$Department of Chemistry, Iowa State University, Ames, Iowa 50011, USA} \\

\end{center}

\newpage

\section{Additional synthetic details and powder X-ray diffraction}

\begin{figure}[!b]
    \centering
    \includegraphics[width=\linewidth]{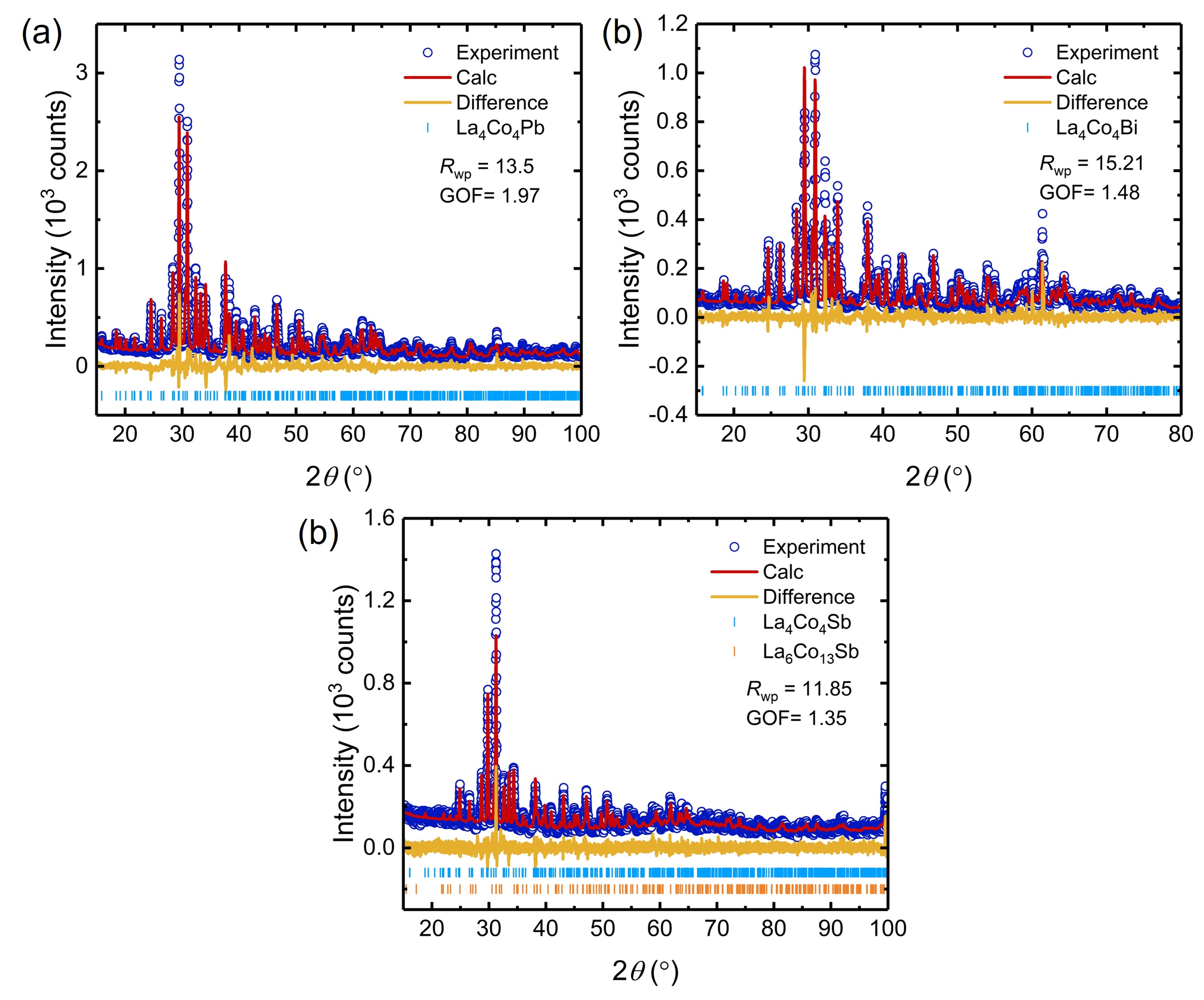}
    \caption[]{Powder X-ray diffraction patterns and Rietveld refinements for (a) La$_4$Co$_4$Pb, (b) La$_4$Co$_4$Bi, and (c) La$_4$Co$_4$Sb.}
    \label{pxrd}
\end{figure}

\begin{figure}[!t]
    \centering
    \includegraphics[width=0.7\linewidth]{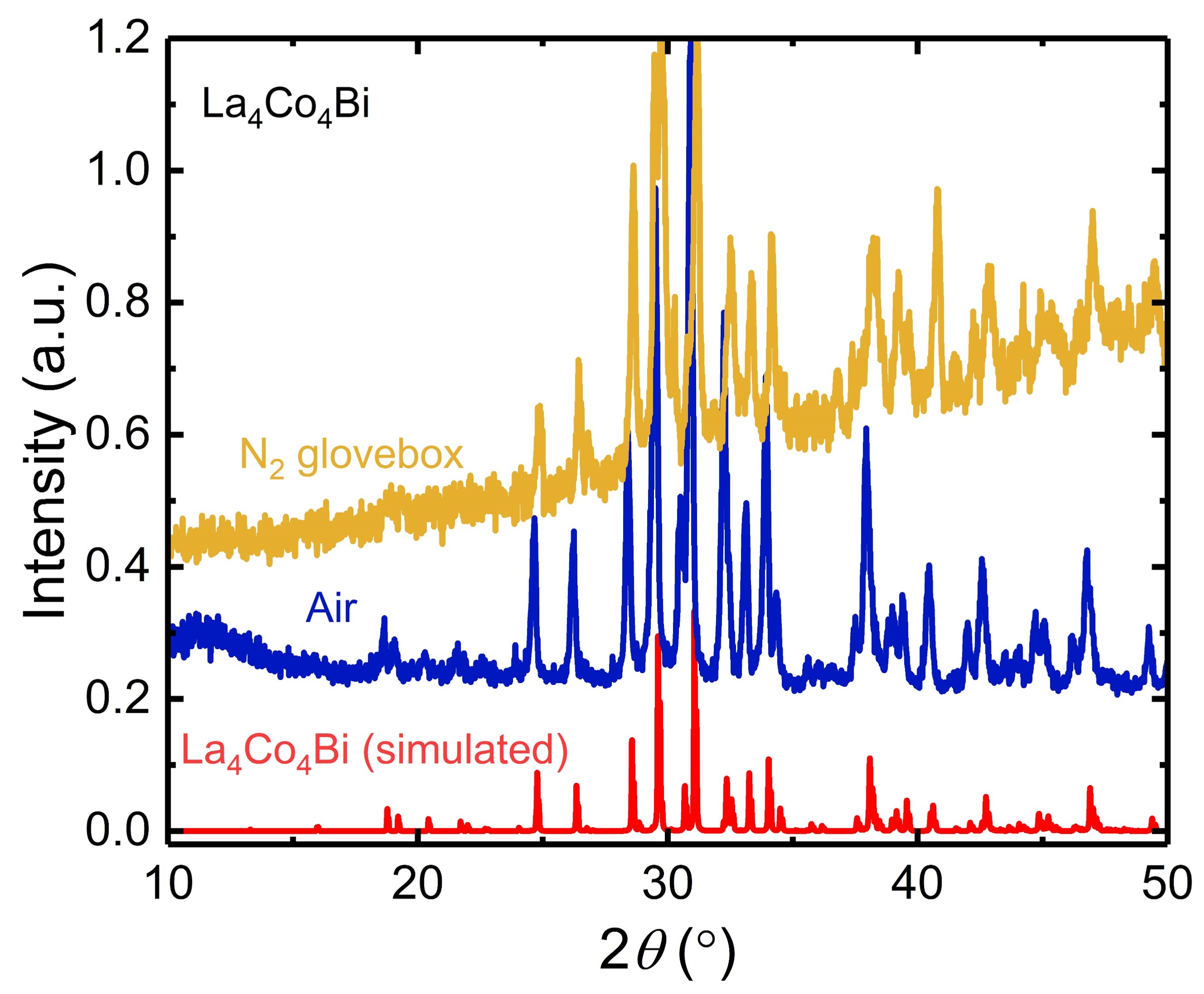}
    \caption[]{Powder X-ray diffraction patterns for a La$_4$Co$_4$Bi sample measured in air (blue) and in a nitrogen filled glovebox (gold). The red curve shows the pattern theoretically expected for La$_4$Co$_4$Bi. The diffractometer in the nitrogen glovebox does not have a fluorescence filter, which gives the gold data a background that increases steadily with angle.}
    \label{pxrdN2}
\end{figure}

Powder X-ray diffraction (PXRD) patterns obtained on the blade-like crystals discussed in the synthetic details section of the main text are presented in Figure \ref{pxrd}. Initially, we were unable to match the patterns with any known phases in the La-Co-\textit{X} phase spaces, suggesting the presence of previously unknown materials. After using single crystal X-ray diffraction to solve the crystal structures, we used the \textit{Pbam} La$_4$Co$_4$\textit{X} structural models to refine the powder patterns with the Rietveld method. The final refinement parameters are in the range R$_{wp}$ = 11-15 and GOF $\approx$ 1.5 for all samples, suggesting the structural models determined from the single crystal data capture the experimental PXRD patterns reasonably well.

Because the surfaces of the crystals were found to darken quickly (within minutes) with exposure to air, we also measured powder X-ray diffraction patterns using a different diffractometer (Rigaku Miniflex I) that is contained inside a nitrogen filled glovebox. Figure \ref{pxrdN2} compares PXRD patterns for La$_4$Co$_4$Bi samples prepared and measured in the glovebox (gold curve) and in air (blue curve). The Figure also shows the theoretical pattern expected for La$_4$Co$_4$Bi. Both experimental datasets are in reasonable agreement with each other and agree well with the theoretical pattern. These results suggest that despite quickly darkening with air exposure, the bulk of the La$_4$Co$_4$\textit{X} samples does not degrade substantially over the timescale of the PXRD data collection ($\approx$ 1 hour).

Having discovered three new compounds, La$_4$Co$_4$\textit{X} where \textit{X} = Pb, Bi, and Sb, we suspected an isostructural La$_4$Co$_4$Sn may also exist. Initially, we attempted to grow a hypothetical La$_4$Co$_4$Sn from solution in an analogous manner to the procedure used for the other La$_4$Co$_4$\textit{X} outlined in the main text. These attempts to grow La$_4$Co$_4$Sn from solution produced mixtures of known binary or ternary phases within the La-Co-Sn phase space. Therefore, we also attempted to prepare La$_4$Co$_4$Sn using direct combination via arc melting. Here, we weighed $\approx$ 1 g total of elemental La, Co, and Sn in a 4:4:1 molar ratio and melted the elements using an arc melter. During arc melting, the button was flipped and remelted several times to improve homogeneity. After arc melting, the La-Co-Sn button was placed in a Ta crucible and vacuum sealed in a fused silica ampule. The ampule was then annealed at 800 C for 3 days, after which the furnace was turned off and naturally cooled to room temperature. Several pieces of the button were then broken off, ground to a fine powder, and analyzed with powder X-ray diffraction. 

\begin{figure*}[!t]
    \centering
    \includegraphics[width=0.7\linewidth]{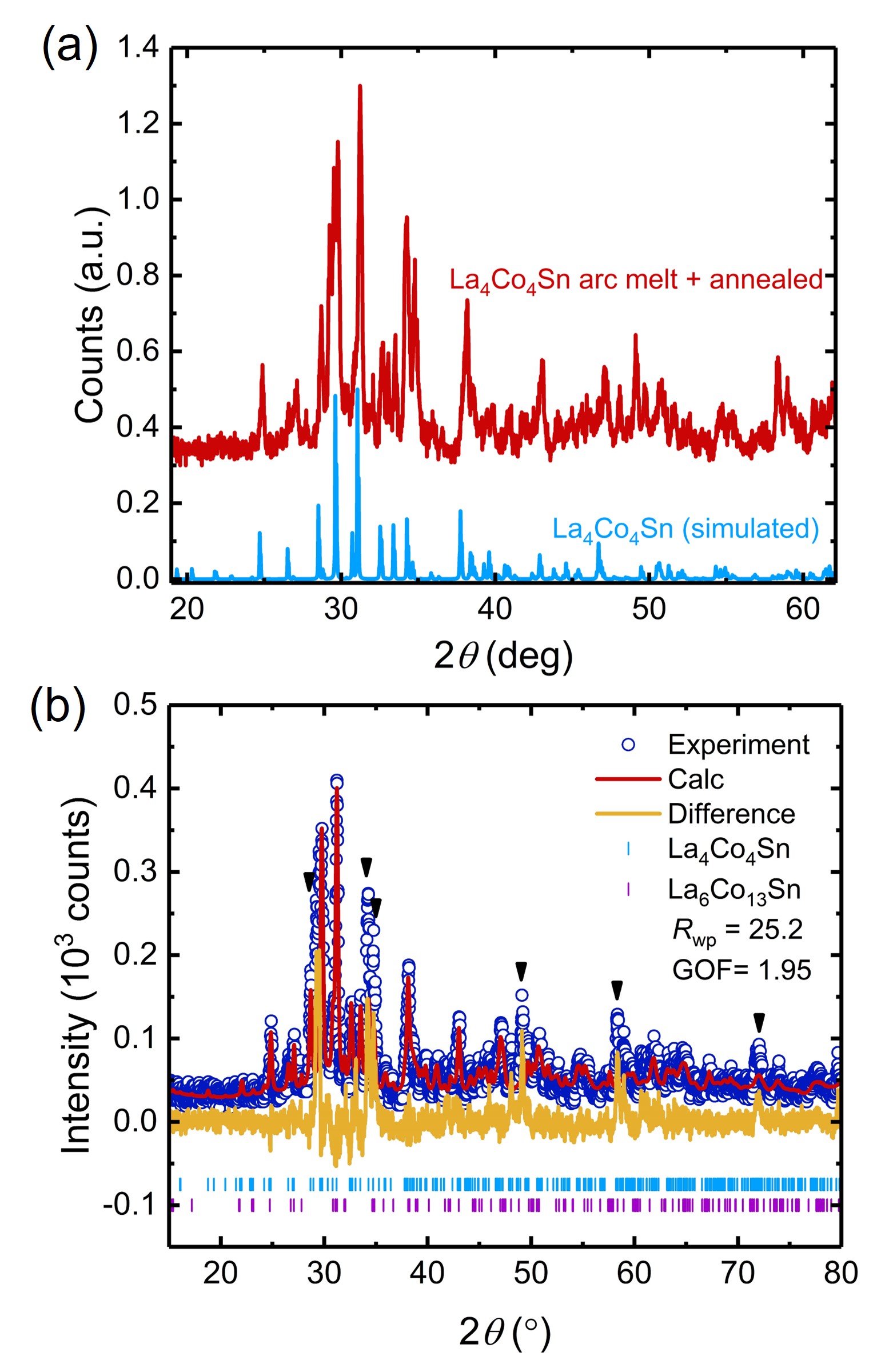}
    \caption[]{Powder X-ray diffraction pattern from a arc-melted and annealed pellet of nominal composition La$_4$Co$_4$Sn. The experimental pxrd pattern is shown in red and the blue lines give the theoretical pattern calculated for a hypothetical La$_4$Co$_4$Sn using the \textit{Pbam} structural model determined from the single crystal refinements of La$_4$Co$_4$\textit{X}. (b) Rietveld refinement of the powder pattern in (a). The primary phases are La$_4$Co$_4$Sn and La$_6$Co$_{13}$Sn. The black arrows mark reflections we were not able to confidently assign.}
    \label{La-Co-Sn_pxrd}
\end{figure*}

The PXRD pattern obtained on the arc-melted and annealed La-Co-Sn button is shown in Figure \ref{La-Co-Sn_pxrd}a and is compared with the theoretical pattern anticipated for a hypothetical La$_4$Co$_4$Sn compound that is isostructural to the \textit{X} = Pb, Bi, and Sb based La$_4$Co$_4$\textit{X} materials. Whereas the pxrd pattern shown in Figure \ref{La-Co-Sn_pxrd} clearly suggests a mixture of phases are present in the final button, many of the peaks match well with those anticipated for La$_4$Co$_4$Sn. Figure \ref{La-Co-Sn_pxrd}b shows a Rietveld refinement of the same powder pattern, and we find the majority of the reflections are consistent with a mixture of primarily La$_4$Co$_4$Sn and La$_6$Co$_{13}$Sn. There are also several strong peaks (marked with black arrows) that we were unable to confidently assign to La-Co-Sn binary or ternary compounds. Despite being unable here to grow single crystals suitable for further analysis or phase-pure powder samples, the the powder patterns in Figure \ref{La-Co-Sn_pxrd} show strong evidence that La$_4$Co$_4$Sn is also a stable phase. Future work to isolate single crystals is highly desirable, as La$_4$Co$_4$Sn would be isoelectronic to La$_4$Co$_4$Pb, and therefore may aid our understanding of the distinct magnetic behavior observed between La$_4$Co$_4$Pb and La$_4$Co$_4$Pn (Pn = Sb, Bi).

\begin{figure}[t!]
    \centering
    \includegraphics[width=0.6\linewidth]{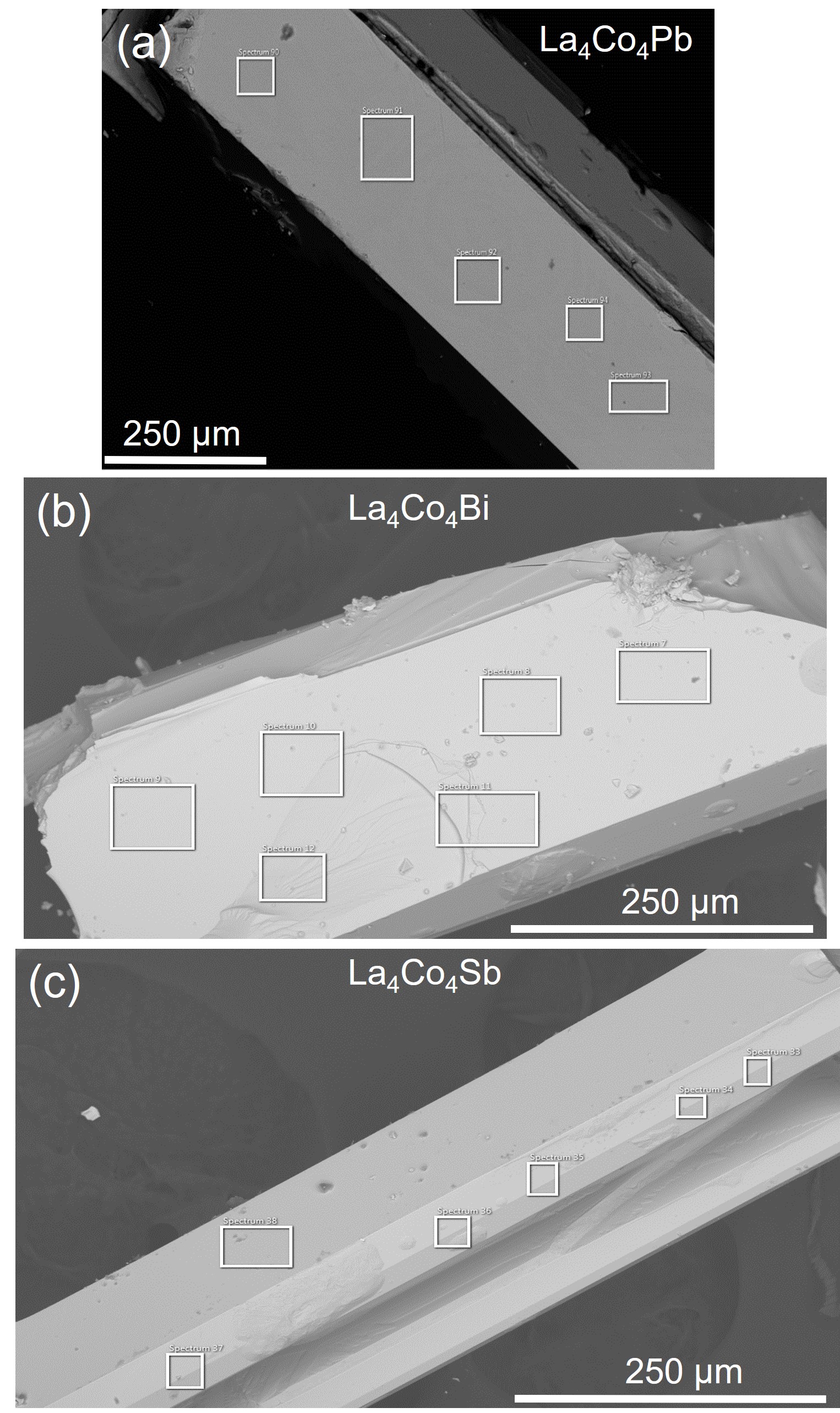}
    \caption[]{Representative SEM images of (a) La$_4$Co$_4$Pb, (b) La$_4$Co$_4$Bi, and (c)La$_4$Co$_4$Sb. The white boxes show the areas from which EDS spectra were obtained.}
    \label{SEM}
\end{figure}

\newpage
\clearpage

\begin{table}[t!]
  \centering
  \caption{EDS data for La$_4$Co$_4$Pb. Data was collected on three separate samples, and the compositions of each are the averages of 4-8 separate spectra collected on each. The uncertainties are the standard deviations of the compositions of each spectra.}
    \begin{tabular}{llrrr}
        &     & \multicolumn{1}{l}{La} & \multicolumn{1}{l}{Co} & \multicolumn{1}{l}{Pb} \\
        \midrule
    \multirow{2}[0]{*}{Sample 1} & Average atomic \% & \multicolumn{1}{l}{45.3(6)} & \multicolumn{1}{l}{43.1(8)} & \multicolumn{1}{l}{11.6(2)} \\
        & Composition & 3.9 & 3.7 & 1 \\
    \multirow{2}[0]{*}{Sample 2} & Average atomic \% & \multicolumn{1}{l}{45.4(1)} & \multicolumn{1}{l}{43.66(8)} & \multicolumn{1}{l}{10.9(2)} \\
        & Composition & 4.1 & 4   & 1 \\
    \multirow{2}[0]{*}{Sample 3} & Average atomic \% & \multicolumn{1}{l}{45(1)} & \multicolumn{1}{l}{42.9(2)} & \multicolumn{1}{l}{11(1)} \\
        & Composition & 4.1 & 3.8 & 1 \\
        \bottomrule
    \end{tabular}%
  \label{EDS_Pb}%
\end{table}%

\begin{table}[t!]
  \centering
  \caption{EDS data for La$_4$Co$_4$Bi. Data was collected on three separate samples, and the compositions of each are the averages of 4-8 separate spectra collected on each. The uncertainties are the standard deviations of the compositions of each spectra.}
    \begin{tabular}{llrrr}
        &     & \multicolumn{1}{l}{La} & \multicolumn{1}{l}{Co} & \multicolumn{1}{l}{Bi} \\
        \midrule
    \multirow{2}[0]{*}{Sample 1} & Average atomic \% & \multicolumn{1}{l}{45.7(3)} & \multicolumn{1}{l}{43.4(7)} & \multicolumn{1}{l}{11(1)} \\
        & Composition & 4.2 & 4   & 1 \\
    \multirow{2}[0]{*}{Sample 2} & Average atomic \% & \multicolumn{1}{l}{45.90(6)} & \multicolumn{1}{l}{43.8(1)} & \multicolumn{1}{l}{10.3(1)} \\
        & Composition & 4.4 & 4.2 & 1 \\
    \multirow{2}[0]{*}{Sample 3} & Average atomic \% & \multicolumn{1}{l}{45.6(4)} & \multicolumn{1}{l}{43(2)} & \multicolumn{1}{l}{11(2)} \\
        & Composition & 4.1 & 3.9 & 1 \\
        \bottomrule
    \end{tabular}%
  \label{EDS_Bi}%
\end{table}%

\begin{table}[t!]
  \centering
  \caption{EDS data for La$_4$Co$_4$Sb. Data was collected on three separate samples, and the compositions of each are the averages of 4-8 separates spectra collected on each. The uncertainties are the standard deviations of the compositions of each spectra.}
    \begin{tabular}{llrrr}
        &     & \multicolumn{1}{l}{La} & \multicolumn{1}{l}{Co} & \multicolumn{1}{l}{Sb} \\
        \midrule
    \multirow{2}[0]{*}{Sample 1} & Average atomic \% & \multicolumn{1}{l}{38(1)} & \multicolumn{1}{l}{37(1)} & \multicolumn{1}{l}{9.1(7)} \\
        & Composition & 4.2 & 4.1 & 1 \\
    \multirow{2}[0]{*}{Sample 2} & Average atomic \% & \multicolumn{1}{l}{34.5(3)} & \multicolumn{1}{l}{33.1(5)} & \multicolumn{1}{l}{8.1(5)} \\
        & Composition & 4.2 & 4.1 & 1 \\
    \multirow{2}[0]{*}{Sample 3} & Average atomic \% & \multicolumn{1}{l}{37.8(3)} & \multicolumn{1}{l}{36.9(3)} & \multicolumn{1}{l}{9.5(1)} \\
        & Composition & 4   & 3.9 & 1 \\
        \bottomrule
    \end{tabular}%
  \label{EDS_Sb}%
\end{table}%

\clearpage
\newpage

\section{Elemental analysis}

We used energy dispersive X-ray spectroscopy to determine the composition of the blade-like crystals. Figures \ref{SEM}a-c show scanning electron microscopy (SEM) images of representative samples, and Tables \ref{EDS_Pb}-\ref{EDS_Sb} list the compositions determined from 4-8 EDS spectra obtained on three separate crystals of each La$_4$Co$_4$\textit{X} compound. As noted in the main text, the crystals quickly darken when exposed to air, and the EDS spectra for each spot always showed a strong peak attributable to oxygen. The intensity of the oxygen peak was not consistent between different samples, suggesting it may be extrinsic. This is supported by the fact that the crystal growth took place in a Ta container sealed under vacuum. Therefore, we assume the oxygen peak represents surface oxidation that occurs when transferring the sample from the glovebox to the SEM chamber, and here we only compare the relative atomic percentages of the elements La, Co, and \textit{X} = Pb, Bi, or Sb. With this assumption, we find that all samples have compositions of approximately La$_4$Co$_4$\textit{X}, in agreement with the compositions determined by refining the single crystal X-ray diffraction data. Measurements of multiple spots on each sample show the composition is reasonably consistent from point to point, indicating good homogeneity. 

\section{Room temperature magnetization isotherms}

\begin{figure}[!t]
    \centering
    \includegraphics[width=0.75\linewidth]{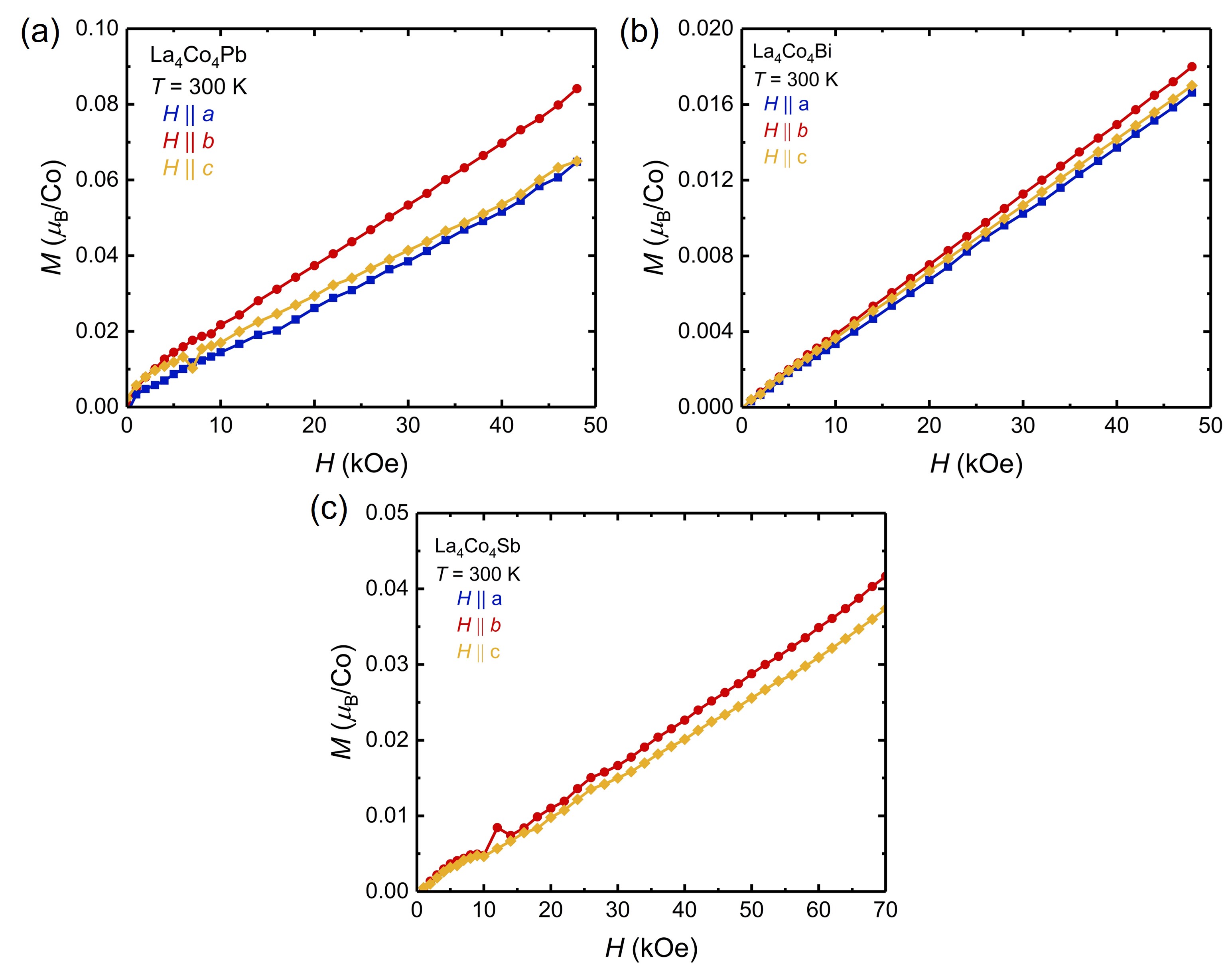}
    \caption[]{Magnetization isotherms measured at 300 K for (a) La$_4$Co$_4$Pb, (b) La$_4$Co$_4$Bi, and (c) La$_4$Co$_4$Sb.}
    \label{MH_300K}
\end{figure}

Figure \ref{MH_300K} shows magnetization vs field isotherms for the La$_4$Co$_4$\textit{X} materials measured at 300 K. All samples show nearly linear behavior at most fields, consistent with a bulk paramagnetic response above the Neel temperatures. In La$_4$Co$_4$Pb and La$_4$Co$_4$Sb, there is also a low-field region of non-linear behavior, indicating these samples contain a very small fraction of a ferromagnetic impurity whose Curie temperature is greater than 300 K. A likely candidate for the second phase is La$_6$Co$_{13}$\textit{X}, which, as we discuss in the synthetic details section of the main text, was present as a second phase in many of our crystal growth attempts. As the La$_6$Co$_{13}$\textit{X} are all reported to be ferromagnetic with Curie temperatures above 400 K,$^{1,2}$ it is likely that small quantities of La$_6$Co$_{13}$\textit{X} may have formed out of droplets of residual flux left on the La$_4$Co$_4$\textit{X} samples, giving the very small ferromagnetic signal observed in Figures \ref{MH_300K}a and \ref{MH_300K}c.

\section{Curie Weiss Fits}

\begin{figure*}[!t]
    \centering
    \includegraphics[width=0.9
\linewidth]{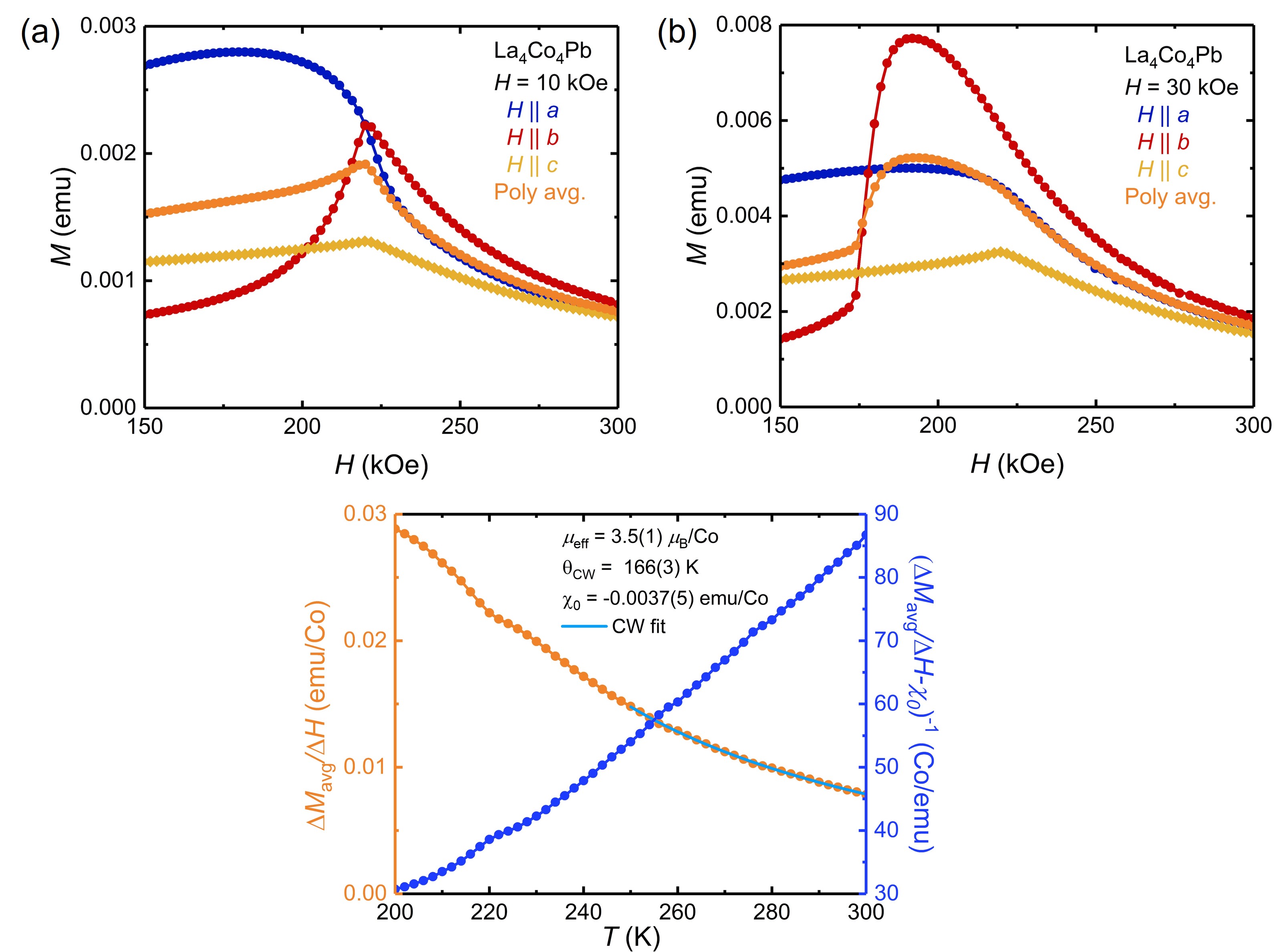}
    \caption[]{Anisotropic temperature dependent magnetization of La$_4$Co$_4$Pb measured at (a) 10 kOe and (b) \textit{H} = 30 kOe fields. We only measured the data between 150-300 K. The orange points show the polycrystalline average of the anisotropic data. (c) Honda-Owen analysis and Curie Weiss fits using the polycrystalline averages of the data in (a) and (b). The left axis (orange) gives the temperature dependence of $\Delta$\textit{M$_\text{avg}$}/$\Delta$\textit{H} and the right axis (blue) ($\Delta$\textit{M$_\text{avg}$}/$\Delta$\textit{H}-$\chi_0$)$^{-1}$. Because the magnetization becomes strongly field-dependent near and below \textit{T}$_N$ = 220 K, we only show the data in (c) between 200-300K.}
    \label{HondaOwen}
\end{figure*}

To determine the effective magnetic moments ($\mu_{\text{eff}}$) in the high temperature paramagnetic state, we fit the magnetization data to a modified Curie Weiss law,

\begin{center}

$\chi^{-1}$ = $\frac{\textit{C}}{\textit{T}-\theta_{\text{CW}}}-\chi_0$

\end{center}

\noindent where \textit{C} is the Curie constant, $\theta_{\text{CW}}$ the Weiss temperature, and $\chi_0$ is a temperature independent term. Because the temperature dependence of \textit{M/H} is not Curie-Weiss like in La$_4$Co$_4$Bi and La$_4$Co$_4$Bi, we only analyzed La$_4$Co$_4$Pb. We used a Honda-Owen method$^{3,4}$ to account for the small ferromagnetic impurity suggested by the 300 K \textit{M}(\textit{H}) isotherms in Figure \ref{MH_300K}. To do so, we measured the magnetization as a function of temperature at two fields, \textit{H} = 10 kOe and 30 kOe, each of which is above the field needed to saturate the ferromagnetic impurity. The anisotropic magnetization curves are shown in Figures \ref{HondaOwen}a and \ref{HondaOwen}b. Using the polycrystalline average of the anisotropic data, where \textit{M}$_{\text{avg}}$/\textit{H} = 1/3(\textit{M}$_{\text{a}}$ + \textit{M}$_{\text{b}}$ + \textit{M}$_{\text{c}}$)/\textit{H}, we can then estimate the magnetic susceptibility at each temperature as $\chi$ $\approx$ $\Delta$\textit{M}/$\Delta$\textit{H}, where $\Delta$\textit{M} = \textit{M}(30 kOe)-\textit{M}(10 kOe) and $\Delta$\textit{H} = 30 kOe - 10 kOe = 20 kOe. We note this data was measured on a different sample as that shown in Figure \ref{MH_300K} and in the main text. In this case, six different La$_4$Co$_4$Pb crystals were co-aligned and glued to a kel-f disc. The magnetic data are in good agreement with those discussed previously, indicating good consistency between different crystals.

Figure \ref{HondaOwen}c shows the temperature dependence of $\Delta$\textit{M}/$\Delta$\textit{H} (left axis, in red) for La$_4$Co$_4$Pb between 200-300 K. By fitting $\Delta$\textit{M}/$\Delta$\textit{H} to the modified Curie Weiss law over 250-300 K, we estimate $\mu_{\text{eff}}$ = 3.5(1) $\mu_{\text{B}}$/Co, the Weiss temperature $\theta_{\text{CW}}$ = 166(3) K, and $\chi_0$ = -0.0037(5) emu/Co. The right axis shows that when using the fitted $\chi_0$, ($\Delta$\textit{M}$_{\text{avg}}$/$\Delta$\textit{H}-$\chi_0$)$^{-1}$ is effectively linear between 230-300 K, consistent with paramagnetic Curie-Weiss like behavior. The 3.5(1) $\mu_{\text{B}}$ moment slightly reduced from, but in reasonable agreement with the 3.87 $\mu_{\text{B}}$ theoretically expected for Co$^{2+}$ local moments. Somewhat unusually, the positive, 166 K, Weiss temperature suggests ferromagnetic fluctuations in the paramagnetic regime, which differs from the observed antiferromagnetic order.

\clearpage
\newpage

\section{Crystallographic tables}

\begin{table}[htbp]
  \centering
  \caption{Atomic positions, site occupancy, and isotropic thermal displacement parameters for La$_4$Co$_4$Pb}
    \begin{tabular}{llllrrl}
    \toprule
    Atom & Site & x   & y   & \multicolumn{1}{l}{z} & \multicolumn{1}{l}{Occ.} & U$_{iso}$ \\
    \midrule
    Pb1 & 4g  & 0.69745(2) & 0.34464(5) & 1   & 1   & 0.01127(12) \\
    La1 & 4g  & 0.42322(2) & 0.90884(7) & 0   & 1   & 0.01157(14) \\
    La2 & 4h  & 0.41837(3) & 0.56508(8) & -0.5 & 1   & 0.01264(14) \\
    La3 & 4h  & 0.72406(2) & 1.05838(7) & 0.5 & 1   & 0.01159(14) \\
    La4 & 4g  & 0.65394(3) & 0.72602(7) & 1   & 1   & 0.01277(14) \\
    Co1 & 8i  & 0.52799(4) & 0.74384(11) & \multicolumn{1}{l}{0.2502(2)} & 1   & 0.0110(2) \\
    Co2 & 2a  & \multicolumn{1}{r}{0.5} & \multicolumn{1}{r}{0.5} & 0   & 1   & 0.0129(4) \\
    Co3 & 2d  & \multicolumn{1}{r}{0.5} & \multicolumn{1}{r}{1} & 0.5 & 1   & 0.0132(4) \\
    Co4 & 4h  & 0.60904(6) & 0.8890(2) & 0.5 & 1   & 0.0156(3) \\
    \bottomrule
    \end{tabular}%
  \label{coord_La4Co4Pb}%
\end{table}%

\begin{table}[htbp]
  \centering
  \caption{Anisotropic thermal displacement parameters for La$_4$Co$_4$Pb.}
    \begin{tabular}{llllrrl}
    \toprule
        & U$_{11}$ & U$_{22}$ & U$_{33}$ & \multicolumn{1}{l}{U$_{23}$} & \multicolumn{1}{l}{U$_{13}$} & U$_{12}$ \\
        \midrule
    Pb1 & 0.00900(16) & 0.01296(19) & 0.01185(19) & 0   & 0   & 0.00011(11) \\
    La1 & 0.0087(2) & 0.0128(3) & 0.0132(3) & 0   & 0   & -0.00049(19) \\
    La2 & 0.0086(2) & 0.0164(3) & 0.0129(3) & 0   & 0   & 0.00086(19) \\
    La3 & 0.0098(2) & 0.0139(3) & 0.0110(3) & 0   & 0   & 0.00047(19) \\
    La4 & 0.0099(2) & 0.0137(3) & 0.0148(3) & 0   & 0   & 0.0011(2) \\
    Co1 & 0.0110(4) & 0.0132(4) & 0.0088(4) & \multicolumn{1}{l}{-0.0007(4)} & \multicolumn{1}{l}{-0.0004(3)} & -0.0013(3) \\
    Co2 & 0.0144(8) & 0.0142(9) & 0.0101(9) & 0   & 0   & -0.0005(7) \\
    Co3 & 0.0147(9) & 0.0140(9) & 0.0107(9) & 0   & 0   & 0.0014(7) \\
    Co4 & 0.0101(6) & 0.0239(8) & 0.0127(7) & 0   & 0   & -0.0053(5) \\
    \bottomrule
    \end{tabular}%
  \label{anis_La4Co4Pb}%
\end{table}%

\clearpage
\newpage

\begin{table}[htbp]
  \centering
  \caption{Atomic positions, site occupancy, and isotropic thermal displacement parameters for La$_4$Co$_4$Bi.}
    \begin{tabular}{llllrrl}
    \toprule
    Atom & Site & x   & y   & \multicolumn{1}{l}{z} & \multicolumn{1}{l}{Occ.} & U$_{iso}$ \\
    \midrule
    Bi1 & 4g  & 0.69895(2) & 0.34009(2) & 1   & 1   & 0.01013(5) \\
    La1 & 4g  & 0.42436(2) & 0.90820(4) & 0   & 1   & 0.01034(6) \\
    La2 & 4h  & 0.42054(2) & 0.56773(4) & -0.5 & 1   & 0.01220(6) \\
    La3 & 4h  & 0.72303(2) & 1.05278(4) & 0.5 & 1   & 0.01033(6) \\
    La4 & 4g  & 0.65040(2) & 0.72255(4) & 1   & 1   & 0.01188(6) \\
    Co1 & 8i  & 0.52913(3) & 0.74339(6) & \multicolumn{1}{l}{0.25066(10)} & 1   & 0.01030(9) \\
    Co2 & 2a  & \multicolumn{1}{r}{0.5} & \multicolumn{1}{r}{0.5} & 0   & 1   & 0.01064(18) \\
    Co3 & 2d  & \multicolumn{1}{r}{0.5} & \multicolumn{1}{r}{1} & 0.5 & 1   & 0.01172(18) \\
    Co4 & 4h  & 0.60904(4) & 0.89720(10) & 0.5 & 1   & 0.01409(14) \\
    \bottomrule
    \end{tabular}%
  \label{coord_La4Co4Bi}%
\end{table}%

\begin{table}[htbp]
  \centering
  \caption{Anisotropic thermal displacement parameters for La$_4$Co$_4$Bi.}
    \begin{tabular}{llllrrl}
    \toprule
    Atom & U$_{11}$ & U$_{22}$ & U$_{33}$ & \multicolumn{1}{l}{U23} & \multicolumn{1}{l}{U$_{13}$} & U$_{12}$ \\
    \midrule
    Bi1 & 0.01035(8) & 0.00898(7) & 0.01106(8) & 0   & 0   & -0.00016(5) \\
    La1 & 0.00916(13) & 0.00960(12) & 0.01227(12) & 0   & 0   & -0.00066(9) \\
    La2 & 0.01018(14) & 0.01385(13) & 0.01256(13) & 0   & 0   & 0.00102(10) \\
    La3 & 0.01124(14) & 0.00978(12) & 0.00997(11) & 0   & 0   & 0.00071(9) \\
    La4 & 0.01102(14) & 0.00978(12) & 0.01483(13) & 0   & 0   & 0.00079(9) \\
    Co1 & 0.0128(2) & 0.0103(2) & 0.00780(19) & \multicolumn{1}{l}{-0.00055(17)} & \multicolumn{1}{l}{-0.00029(16)} & -0.00111(17) \\
    Co2 & 0.0146(5) & 0.0086(4) & 0.0087(4) & 0   & 0   & -0.0022(3) \\
    Co3 & 0.0154(5) & 0.0105(4) & 0.0093(4) & 0   & 0   & 0.0020(3) \\
    Co4 & 0.0128(4) & 0.0189(4) & 0.0105(3) & 0   & 0   & -0.0047(3) \\
    \bottomrule
    \end{tabular}%
  \label{anis_La4Co4Bi}%
\end{table}%

\clearpage
\newpage

\begin{table}[htbp]
  \centering
  \caption{Atomic positions, site occupancy, and isotropic thermal displacement parameters for La$_4$Co$_4$Sb.}
    \begin{tabular}{llllrrl}
    \toprule
    Atom & Site & x   & y   & \multicolumn{1}{l}{z} & \multicolumn{1}{l}{Occ.} & U$_{iso}$ \\
    \midrule
    Sb1 & 4g  & 0.69886(2) & 0.34177(3) & 1   & 1   & 0.01018(5) \\
    La1 & 4g  & 0.42334(2) & 0.90701(3) & 0   & 1   & 0.01075(5) \\
    La2 & 4h  & 0.41914(2) & 0.56840(3) & -0.5 & 1   & 0.01244(5) \\
    La3 & 4h  & 0.72461(2) & 1.05676(3) & 0.5 & 1   & 0.01048(5) \\
    La4 & 4g  & 0.65232(2) & 0.72283(3) & 1   & 1   & 0.01208(5) \\
    Co1 & 8i  & 0.52957(2) & 0.74366(4) & \multicolumn{1}{l}{0.25057(7)} & 1   & 0.01066(6) \\
    Co2 & 2a  & \multicolumn{1}{r}{0.5} & \multicolumn{1}{r}{0.5} & 0   & 1   & 0.01129(12) \\
    Co3 & 2d  & \multicolumn{1}{r}{0.5} & \multicolumn{1}{r}{1} & 0.5 & 1   & 0.01195(12) \\
    Co4 & 4h  & 0.61032(2) & 0.89670(7) & 0.5 & 1   & 0.01478(10) \\
    \bottomrule
    \end{tabular}%
  \label{coord_La4Co4Sb}%
\end{table}%

\begin{table}[htbp]
  \centering
  \caption{Anisotropic thermal displacement parameters for La$_4$Co$_4$Bi.}
    \begin{tabular}{llllrrl}
    \toprule
    Atom & U$_{11}$ & U$_{22}$ & U$_{33}$ & \multicolumn{1}{l}{U$_{23}$} & \multicolumn{1}{l}{U$_{13}$} & U$_{12}$ \\
    \midrule
    Sb1 & 0.00953(9) & 0.00970(10) & 0.01132(10) & 0   & 0   & -0.00012(7) \\
    La1 & 0.00871(8) & 0.01067(9) & 0.01287(9) & 0   & 0   & -0.00049(6) \\
    La2 & 0.00970(8) & 0.01460(10) & 0.01301(9) & 0   & 0   & 0.00083(6) \\
    La3 & 0.01018(8) & 0.01069(9) & 0.01058(9) & 0   & 0   & 0.00049(6) \\
    La4 & 0.01027(8) & 0.01077(9) & 0.01520(10) & 0   & 0   & 0.00087(6) \\
    Co1 & 0.01246(14) & 0.01105(15) & 0.00846(14) & \multicolumn{1}{l}{-0.00068(11)} & \multicolumn{1}{l}{-0.00030(10)} & -0.00094(10) \\
    Co2 & 0.0151(3) & 0.0102(3) & 0.0086(3) & 0   & 0   & -0.0022(2) \\
    Co3 & 0.0148(3) & 0.0115(3) & 0.0096(3) & 0   & 0   & 0.0024(2) \\
    Co4 & 0.0119(2) & 0.0213(3) & 0.0112(2) & 0   & 0   & -0.00487(17) \\
    \bottomrule
    \end{tabular}%
  \label{anis_La4Co4Sb}%
\end{table}%

\clearpage
\newpage

\section{References}

(1) Weitzer, F.; Leithe-Jasper, A.; Rogl, P.; Hiebl, K.; No{\"e}l, H.; Wiesinger, G.; Steiner, W. Magnetism of (Fe, Co)-based alloys with the La$_6$Co$_{11}$Ga$_3$-type. \textit{Journal of Solid State Chemistry} \textbf{1993}, 104, 368–376.

\vskip 0.5cm

\noindent (2) Guloy, A. M.; Corbett, J. D. Exploration of the interstitial derivatives of La$_5$Pb$_3$ (Mn$_3$Si$_3$-type). \textit{Journal of Solid State Chemistry} \textbf{1994}, 109, 352–358.

\vskip 0.5cm

\noindent (3) Honda, K. Die thermomagnetischen Eigenschaften der Elemente. \textit{Annalen der Physik} \textbf{1910}, 337, 1027–1063.

\vskip 0.5cm

\noindent (4) Owen, M. Magnetochemische Untersuchungen. Die thermomagnetischen Eigenschaften der Elemente. II. \textit{Annalen der Physik} \textbf{1912}, 342, 657–699. S13


\end{singlespace}